\documentclass[aps,prd,reprint,showkeys,noshowpacs,longbibliography,floatfix,amsmath,amssymb,nofootinbib,noshowpacs,superscriptaddress,notitlepage,preprintnumbers]{revtex4-2}

\usepackage[utf8]{inputenc}
\usepackage[english]{babel}
\usepackage{mathtools}
\usepackage{MnSymbol}
\usepackage{multirow}
\usepackage{tabularx}
\newcolumntype{C}{>{\centering\arraybackslash}X}
\newcolumntype{R}{>{\raggedleft\arraybackslash}X}
\newcolumntype{L}{>{\raggedright\arraybackslash}X}

\usepackage{xr-hyper}
\usepackage[pdftex,bookmarks=true]{hyperref}
\hypersetup{
   colorlinks,
   citecolor=blue,
   filecolor=black,
   linkcolor=blue,
   urlcolor=blue
}
\usepackage{cleveref}
\Crefformat{equation}{Eq.~(#2#1#3)}
\Crefmultiformat{equation}{Eqs.~(#2#1#3)}{ and~(#2#1#3)}{, (#2#1#3)}{ and~(#2#1#3)}
\Crefformat{figure}{Fig.~#2#1#3}
\Crefformat{section}{Sec.~#2#1#3}

\allowdisplaybreaks

\def\latmom{2\pi/L}
\newcommand{\bv}[1]{{\mathbf{#1}}}

\begin{document}

\title{Moments and power corrections of longitudinal and transverse proton structure functions from lattice QCD}

\author{M.~Batelaan}
\affiliation{CSSM, Department of Physics, The University of Adelaide, Adelaide SA 5005, Australia}
\author{K.~U.~Can}
\affiliation{CSSM, Department of Physics, The University of Adelaide, Adelaide SA 5005, Australia}
\author{A.~Hannaford-Gunn}
\affiliation{CSSM, Department of Physics, The University of Adelaide, Adelaide SA 5005, Australia}
\author{R.~Horsley}
\affiliation{School of Physics and Astronomy, University of Edinburgh, Edinburgh EH9 3JZ, UK}
\author{Y.~Nakamura}
\affiliation{RIKEN Center for Computational Science, Kobe, Hyogo 650-0047, Japan}
\author{H.~Perlt}
\affiliation{Institut f\"{u}r Theoretische Physik, Universit\"{a}t Leipzig, 04103 Leipzig, Germany}
\author{P.~E.~L.~Rakow}
\affiliation{Theoretical Physics Division, Department of Mathematical Sciences, University of Liverpool, Liverpool L69 3BX, United Kingdom}
\author{G.~Schierholz}
\affiliation{Deutsches Elektronen-Synchrotron DESY, Notkestr. 85, 22607 Hamburg, Germany.}
\author{H.~St\"{u}ben}
\affiliation{Regionales Rechenzentrum, Universit\"{a}t Hamburg, 20146 Hamburg, Germany}
\author{R.~D.~Young}
\affiliation{CSSM, Department of Physics, The University of Adelaide, Adelaide SA 5005, Australia}
\author{J.~M.~Zanotti}
\affiliation{CSSM, Department of Physics, The University of Adelaide, Adelaide SA 5005, Australia}

\collaboration{QCDSF/UKQCD/CSSM Collaborations}
\noaffiliation

\begin{abstract}
	We present a simultaneous extraction of the moments of $F_2$ and $F_L$ structure functions of the proton for a range of photon virtuality, $Q^2$. This is achieved by computing the forward Compton amplitude on the lattice utilizing the second-order Feynman-Hellmann theorem. Our calculations are performed on configurations with two different lattice spacings and volumes, all at the $SU(3)$ symmetric point. We find the moments of $F_{2}$ and $F_{L}$ in good agreement with experiment. Power corrections turn out to be significant. This is the first time the $Q^2$ dependence of the lowest moment of $F_2$ has been quantified. 
\end{abstract}

\keywords{nucleon structure, parton distributions, Feynman Hellmann, Compton amplitude, transverse, longitudinal, structure functions, power corrections, scaling, lattice QCD}
\preprint{ADP-22-27/T1198, DESY-22-145, Liverpool LTH 1307}
\maketitle

\section{Introduction} \label{sec:intro}
Nucleon structure functions are encoded by the differential cross sections for inclusive electron--proton scattering. In terms of the partonic structure of the nucleon, the deep inelastic cross sections are dominated by the transverse structure function, $F_2$, which hence provides the primary constraint on the parton distributions. On the other hand, the longitudinal structure function, $F_L$, provides important information on the QCD structure of the proton. With a perturbatively small and calculable leading-twist component~\cite{ALTARELLI197889}, $F_L$ offers a direct measure of higher-twist effects~\cite{Choi:1993cu}. It also offers sensitivity to the low-$x$ gluon distribution~\cite{Cooper-Sarkar:1987cnv}.

Although the small nature of the longitudinal structure function makes it more challenging to isolate, measurements by HERA~\cite{H1:2010fzx} and Jefferson Lab~\cite{JeffersonLabHallCE94-110:2004nsn,JLab:web} have enabled a direct extraction of several low moments of $F_L$ across a range of $Q^2$~\cite{Monaghan:2012et}. The results reveal a tension with global PDF fits~\cite{Accardi:2011fa,Alekhin:2009ni,Martin:2009iq} at lower $Q^2$ that might indicate non-negligible higher-twist effects or an increased high-$x$ gluon distribution~\cite{Monaghan:2012et}. It is therefore highly desirable to be able to provide first-principles theoretical predictions regarding $F_L$, preferably at intermediate $Q^2$ values where the non-perturbative effects become significant. Furthermore, an improved theoretical constraint on power corrections in the structure functions generally could be particularly beneficial in global PDF analyses~\cite{Alekhin:1999iq,Martin:2003sk,Blumlein:2008kz,Martin:2009iq,Accardi2010,Alekhin:2012ig,Abt2016,HarlandLang2016,Alekhin:2017kpj}.

Lattice QCD simulations of the structure functions conventionally utilise the operator product expansion (OPE) approach. Lattice simulations have been successful in computing the twist-2 contributions, however the higher-twist terms mix with those of lower-twist which gives rise to complications in the renormalisation procedure~\cite{Martinelli:1996pk}. This setback has limited lattice QCD to investigations of the leading-twist contributions~\cite{LIN2018107,Constantinou:2022yye}, with fewer works on twist-3 contributions~\cite{Gockeler:2005vw,Bhattacharya:2020cen,Bhattacharya:2021moj}.

In this work, we present a simultaneous extraction of the low moments of the nucleon structure functions $F_2$ and $F_L$ from the forward Compton amplitude calculated on the lattice. This approach circumvents the operator mixing issues since the amplitude accounts for the mixing and renormalisation and contains all twist contributions. Previous successful calculations of the Compton amplitude, leading to a determination of the moments of the nucleon structure function $F_1$, have been reported in~\cite{PhysRevLett.118.242001,PhysRevD.102.114505}, and recently extended to off-forward kinematics~\cite{Alec:2021lkf}.    

\section{Compton amplitude and moments of structure functions} \label{sec:ca}
In order to access the structure functions, we consider the unpolarised forward Compton tensor,
\begin{align}\label{eq:comptensor}
	\begin{split}
		T_{\mu\nu}(p,q) &= \left( -g_{\mu\nu} + \frac{q_\mu q_\nu}{q^2} \right) \mathcal{F}_1(\omega,Q^2)
		+ \frac{\hat{P}_\mu \hat{P}_\nu}{p \cdot q} \mathcal{F}_2(\omega,Q^2),
	\end{split}
\end{align}
where $q$ ($p$) is the momentum of the virtual photon (nucleon), $\hat{P}_\mu \equiv p_\mu - (p \cdot q) q_\mu/q^2$, $\omega = (2p \cdot q)/Q^2$ and $Q^2 = -q^2$. The Lorentz invariant Compton structure functions $\mathcal{F}_{1,2}$ are related to the physical structure functions $F_{1,2}$ via the optical theorem, $\operatorname{Im}\mathcal{F}_{1,2}(\omega,Q^2) = 2\pi F_{1,2}(x,Q^2)$. Making use of analyticity, crossing symmetry, and the optical theorem, the Compton structure functions satisfy the familiar dispersion relations~\cite{Drechsel:2002ar},
\begin{align} 
	\label{eq:compdisp1}
	\overline{\mathcal{F}}_1(\omega,Q^2) &= 2\omega^2 \int_0^1 dx \frac{2x \, F_1(x,Q^2)}{1-x^2\omega^2-i\epsilon}, \\
	\label{eq:compdisp2}
	\mathcal{F}_2(\omega,Q^2) &= 4\omega \int_{0}^1 dx\, \frac{F_2(x,Q^2)}{1-x^2\omega^2-i\epsilon},
\end{align}
where $\overline{\mathcal{F}}_1(\omega,Q^2) = \mathcal{F}_1(\omega,Q^2)-\mathcal{F}_1(0,Q^2)$. 

The parametrisation of the forward Compton amplitude in terms of $F_1$ and $F_2$ is not unique. Alternatively, we can consider a parametrisation in terms of the transverse, $2xF_1$, and longitudinal, $F_L$, structure functions~\cite{Hand:1963bb,Bodek:1979rx,Drechsel:2002ar,Melnitchouk:2005zr}. The latter is given by~\cite{Bodek:1979rx,Drechsel:2002ar},
\begin{equation}\label{eq:FL_x}
	F_L(x,Q^2) = \left( 1 - \frac{4 M_N^2}{Q^2} x^2 \right) F_2(x,Q^2) - 2xF_1(x,Q^2),
\end{equation} 
which can directly be obtained from the ratio of cross sections~\cite{Bodek:1979rx,Melnitchouk:2005zr}. Here $M_N$ is the mass of the nucleon. As $Q^2 \to \infty$, \Cref{eq:FL_x} reduces to $F_L(x) \to F_2(x) - 2xF_1(x) $, which vanishes in the quark-parton model due to the familiar Callan-Gross relation. In QCD, $F_L$ is $\mathcal{O}(\alpha_s)$ suppressed at leading twist and any power correction may be identified as higher twist.

Writing,
\begin{equation} \label{eq:FL_comp}
	\mathcal{F}_L(\omega, Q^2) = -\mathcal{F}_1(\omega,Q^2) + \left(\frac{\omega}{2} + \frac{2 M_N^2}{\omega Q^2} \right) \mathcal{F}_2(\omega,Q^2),
\end{equation}
we can express $\overline{\mathcal{F}}_L$ by a subtracted dispersion relation in terms of $F_L$,
\begin{align}\label{eq:compdispL}
	\begin{split}
		\overline{\mathcal{F}}_L(\omega,Q^2) &= \frac{8M_N^2}{Q^2} \int_0^1 dx F_2(x,Q^2) \\ 
		&+ 2\omega^2 \int_0^1 dx \frac{F_L(x,Q^2)}{1 - x^2 \omega^2 -i\epsilon},
	\end{split}
\end{align}
where $\overline{\mathcal{F}}_L(\omega,Q^2) = \mathcal{F}_L(\omega,Q^2) + \mathcal{F}_1(0,Q^2)$. 

We isolate the Compton structure functions from the tensor in \Cref{eq:comptensor}. Working in Minkowski space and setting $q_3=p_3=0$ we have
\begin{align} 
	\label{eq:compF1}
	\mathcal{F}_1(\omega, Q^2) &= T_{33}(p,q), \\
 	\label{eq:compF2}
    \mathcal{F}_2(\omega,Q^2) &= \frac{\omega Q^2}{2 E_N^2} \left[ T_{00}(p,q) + T_{33}(p,q) \right].
\end{align}
$\mathcal{F}_L$ is constructed via \Cref{eq:FL_comp}.

Expanding the integrands in \Cref{eq:compdisp1,eq:compdisp2,eq:compdispL} as a geometric series, we express the Compton structure functions as infinite sums over the Mellin moments of the inelastic structure functions,
\begin{align} 
	\label{eq:ope_moments1}
	\overline{\mathcal{F}}_{1,L}(\omega,Q^2)&=\sum_{n=0}^\infty 2\omega^{2n} M^{(1,L)}_{2n}(Q^2), \\
	\label{eq:ope_moments2}
	\mathcal{F}_2(\omega,Q^2)&= \sum_{n=1}^\infty 4\omega^{2n-1} M^{(2)}_{2n}(Q^2),
\end{align}
where $M^{(1)}_{0}(Q^2) = 0$, $2M^{(L)}_{0}(Q^2) = \frac{8M_N^2}{Q^2} M^{(2)}_{2}(Q^2)$,
\begin{align} 
	\label{eq:moments1}
	M^{(1)}_{2n}(Q^2) &= 2\int_0^1 dx\, x^{2n-1} F_1(x,Q^2), \\
	\label{eq:moments2}
	M^{(2,L)}_{2n}(Q^2) &= \int_{0}^1 dx\,x^{2n-2} F_{2,L}(x,Q^2),
\end{align}
for $n > 0$.

For our purposes, it is convenient to express the expansion of $\mathcal{F}_2$ in terms of the independently positive definite moments of $F_1$ and $F_L$,
\begin{align}\label{eq:moments_F2}
	\frac{\mathcal{F}_2(\omega)}{\omega} = \frac{\tau}{\left(1 + \tau \, \omega^2 \right)} \sum_{n=0}^{\infty} 4\omega^{2n} \left[ M_{2n}^{(1)} + M_{2n}^{(L)} \right],
\end{align}
where $\tau=Q^2 / 4 M_N^2$. The intercept at $\omega=0$ is proportional to the lowest moment of $F_2$, i.e. $M_2^{(2)}(Q^2)$. Higher moments are given by the appropriate combinations of the moments of $F_1$ and $F_L$. 

In the following discussion, we provide the details of our procedure for extracting the moments directly from the Compton amplitude obtained in a lattice simulation.

\section{The Feynman-Hellmann approach} \label{sec:fh}
The novel idea is to compute the Compton amplitude by means of the second-order Feynman-Hellmann theorem as derived and described in detail in~\cite{PhysRevD.102.114505}. Here we summarise the procedure relevant to this work. We perturb the fermion action by the vector current,
\begin{equation}\label{eq:fh_perturb}
    S(\lambda) = S + \lambda \int d^3z (e^{i \bv{q} \cdot \bv{z}} + e^{-i \bv{q} \cdot \bv{z}}) \mathcal{J}_{\mu}(z),
\end{equation}
where $\lambda$ is the strength of the coupling between the quarks and the external field, $\mathcal{J}_{\mu}(z) = Z_V \bar{q}(z) \gamma_\mu q(z)$ is the electromagnetic current coupling to the quarks, $\bv{q}$ is the external momentum inserted by the current and $Z_V$ is the renormalisation constant for the local electromagnetic current, which has been determined in Ref~\cite{Constantinou:2014fka}. The perturbation is introduced on the valence quarks only, hence only quark-line connected contributions are taken into account in this work. For the perturbation of valence and sea quarks see~\cite{Chambers2015}.

We consider $q_3=p_3=0$ and current components $\mathcal{J}_{0}$ and $\mathcal{J}_{3}$, enabling us to compute $T_{00}$ and $T_{33}$. These are then given by the second order energy shift~\cite{PhysRevD.102.114505},
\begin{equation} \label{eq:secondorder_fh}
    \left. \frac{\partial^2 E_{N_\lambda}(\bv{p})}{\partial \lambda^2} \right|_{\lambda=0} = - \frac{T_{\mu\mu}(p,q) + T_{\mu\mu}(p,-q)}{2 E_{N}(\bv{p})},
\end{equation}
where $T_{\mu\nu}$ is the Compton tensor defined in \Cref{eq:comptensor}, $q=(0,\bv{q})$ is the external momentum encoded by \Cref{eq:fh_perturb}, and $E_{N_\lambda}(\bv{p})$ is the nucleon energy at momentum $\bv{p}$ in the presence of a background field of strength $\lambda$. This expression is the principal relation that we use to access the Compton amplitude and hence the Compton structure functions given in \Cref{eq:compF1,eq:compF2}.

\section{Simulation and analysis} \label{sec:simu}
Our lattice simulations are carried out on QCDSF/UKQCD-generated $2+1$-flavour gauge configurations. We utilise two ensembles with volumes $V = [32^3 \times 64, 48^3 \times 96]$, and couplings $\beta = [5.50, 5.65]$ corresponding to lattice spacings $a = [0.074(2), 0.068(3)] \, {\rm fm}$ respectively. The quark masses are tuned to the $SU(3)$ symmetric point where the masses of all three quark flavours are set to approximately the physical flavour-singlet mass, $\overline{m} = (2 m_s + m_l)/3$~\cite{Bietenholz:2010jr,Bietenholz:2011qq}, yielding $m_\pi \approx [470, 420] \, {\rm MeV}$. We perform up to $\mathcal{O}(10^4)$ and $\mathcal{O}(10^3)$ measurements by employing up to six and three sources on the $32^3 \times 64$ and $48^3 \times 96$ ensembles of size $1764$ and $537$ configurations, respectively.

We follow the procedure laid out in Ref.~\cite{PhysRevD.102.114505} to calculate the energy shifts and extract the Compton amplitude. The calculations are done for several values of $\bv{q}$. Multiple values of $\omega$ are accessed by varying the nucleon momentum $\bv{p}$ for a fixed $\bv{q}$. A list of $\omega$ values used in the analysis is provided in \Cref{app:enshfit}. 

By attaching the current selectively to the $u$ and $d$ quarks, respectively, we obtain the flavour diagonal contributions $uu$ and $dd$ corresponding to a handbag diagram at leading twist, and the mixed-flavour piece, $ud$, which is purely higher-twist, corresponding to a cat's ears diagram~\footnote{Note that we are mentioning the leading-twist diagrams for the clarity of the discussion. In reality, the Compton amplitude includes all twist contributions.}. We construct the ratios,
\begin{align}
	\label{eq:ratio_fd}
	\begin{split}
		\mathcal{R}^{qq}_{\lambda}(\bv{p}, t) &\equiv \frac{G^{(2)}_{+\lambda}(\bv{p}, t) G^{(2)}_{-\lambda}(\bv{p}, t)}{\left( G^{(2)}(\bv{p}, t) \right)^2} \\
		& \xrightarrow{t \gg 0} A_{\lambda}^{qq} e^{-2\Delta E^{qq}_{N_\lambda}(\bv{p}) \, t},
	\end{split} \\
	\label{eq:ratio_fm}
	\begin{split}
		\mathcal{R}^{q q^\prime}_{\lambda}(\bv{p},t) &\equiv \frac{G^{(2)}_{+\lambda,+\lambda}(\bv{p},t) G^{(2)}_{-\lambda,-\lambda}(\bv{p},t)}{G^{(2)}_{+\lambda,-\lambda}(\bv{p},t) G^{(2)}_{-\lambda,+\lambda}(\bv{p},t)} \\
		&\xrightarrow{t \gg 0} A_\lambda^{q q^\prime} e^{-4\Delta E^{q q^\prime}_{N_\lambda}(\bv{p}) \, t},
	\end{split}
\end{align} 
in order to extract the second-order energy shifts for the flavour-diagonal ($qq = uu$, $dd$) and mixed-flavour ($q q^\prime = ud$) pieces, respectively. Here, $G^{(2)}_{\lambda}$ denote the perturbed two-point correlation functions in the presence of the external field with the coupling strength $\lambda$. In order to calculate the $ud$ piece as in \Cref{eq:ratio_fm}, we need to consider the interference of two currents. Therefore we compute the perturbed correlators, $G^{(2)}_{\lambda_1, \lambda_2}$, by including an additional current term in \Cref{eq:fh_perturb} with the same coupling strength in magnitude, $|\lambda_1| = |\lambda_2| = |\lambda|$ in close analogy to the off-forward case~\cite{Alec:2021lkf}. These ratios isolate the energy shifts ($\Delta E_{N_\lambda}^{(qq, q q^\prime)}(\bv{p})$) only at even orders of $\lambda$.

We proceed with established spectroscopy methods to extract the energy shifts from the ratios defined in \Cref{eq:ratio_fd,eq:ratio_fm}. Fit windows are determined following a covariance-matrix based $\chi^2$ analysis. We perform correlated, one-exponential fits to a range of fit windows that contain at least four time slices and pick the one with the best $\chi^2$ per degree of freedom, i.e. $\chi^2_{dof} \sim 1.0$. The majority of the chosen fit windows satisfy this criteria. Any systematic error due to the choice of fit windows could be accounted for by a weighted-averaging method~\cite{NPLQCD:2020ozd,Batelaan:2022fdq}. At our current precision, we find the energy shifts that are extracted via both methods to be in good agreement. Therefore, we continue with simple one-exponential fits.

We typically compute the energy shifts $\Delta E_{N_\lambda}(\bv{p})$, for two $|\lambda|$ values and perform polynomial fits of the form,
\begin{equation} \label{eq:lamfit}
	\Delta E_{N_\lambda}(\bv{p}) = \lambda^2 \left. \frac{\partial^2 E_{N_\lambda}(\bv{p})}{\partial \lambda^2} \right|_{\lambda = \bv{0}} + \mathcal{O}(\lambda^4), 
\end{equation}
to determine the Compton amplitude (see Ref.~\cite{PhysRevD.102.114505}). Choosing $|\lambda| = \mathcal{O}(10^{-2})$, higher order $\mathcal{O}(\lambda^4)$ terms are heavily suppressed. Effective mass plot analogues for the correlator ratios and their corresponding $\lambda$-fits are shown in \Cref{app:enshfit}.

The $\omega$ dependence of the Compton structure functions is mapped by extracting the amplitude for each pair of ($\bv{q}$,$\bv{p}$). Subsequently, extraction of the moments from the Compton structure functions follows the methodology described in~\cite{PhysRevD.102.114505}. A simultaneous fit of $\overline{\mathcal{F}}_1$ (\Cref{eq:ope_moments1}) and $\mathcal{F}_2/\omega$ (\Cref{eq:moments_F2}) is performed in a Bayesian framework to determine the first few Mellin moments of the structure functions. We truncate both series at $n=4$ (inclusive) when determining the moments. These moments are enforced to be positive definite and monotonically decreasing. Note that the positivity bound does not hold for the $ud$ contributions but they are constrained by $\left| M_{2n}^{ud}(Q^2) \right|^2 \le 4 M_{2n}^{uu}(Q^2) M_{2n}^{dd}(Q^2)$, since the total inclusive cross section (hence each moment) is positive for any value of the quark charges and at all kinematics. The sequences of individual $uu$, $dd$ or $ud$ moments are selected according to the standard probability distribution, $\operatorname{exp}(-\chi^2/2)$, where
\begin{equation}
	\chi^2 = \sum_{\mathcal{F}} \sum_{i} \frac{\left[ \mathcal{F}^\text{model}_i - \mathcal{F}^\text{obs}(\omega_i) \right]^2}{\sigma^2}
\end{equation}
is the $\chi^2$ function with $\sigma^2$ the diagonal elements of the full covariance matrix. Here, $\mathcal{F}$ stands for $\overline{\mathcal{F}}_1$ and $\mathcal{F}_2$, and the index $i$ runs through all the $\omega$ values and flavour-diagonal and mixed-flavour pieces. A posterior distribution is obtained for each moment on each bootstrap sample. Then, we resample from these distributions to form a single posterior distribution for each moment to account for the correlations between the data points. Representative posterior distributions for the lowest moments are shown in~\Cref{app:bayes}.

\section{Results} \label{sec:res}
\begin{figure}[ht]
    \centering
    \includegraphics[width=.49\textwidth]{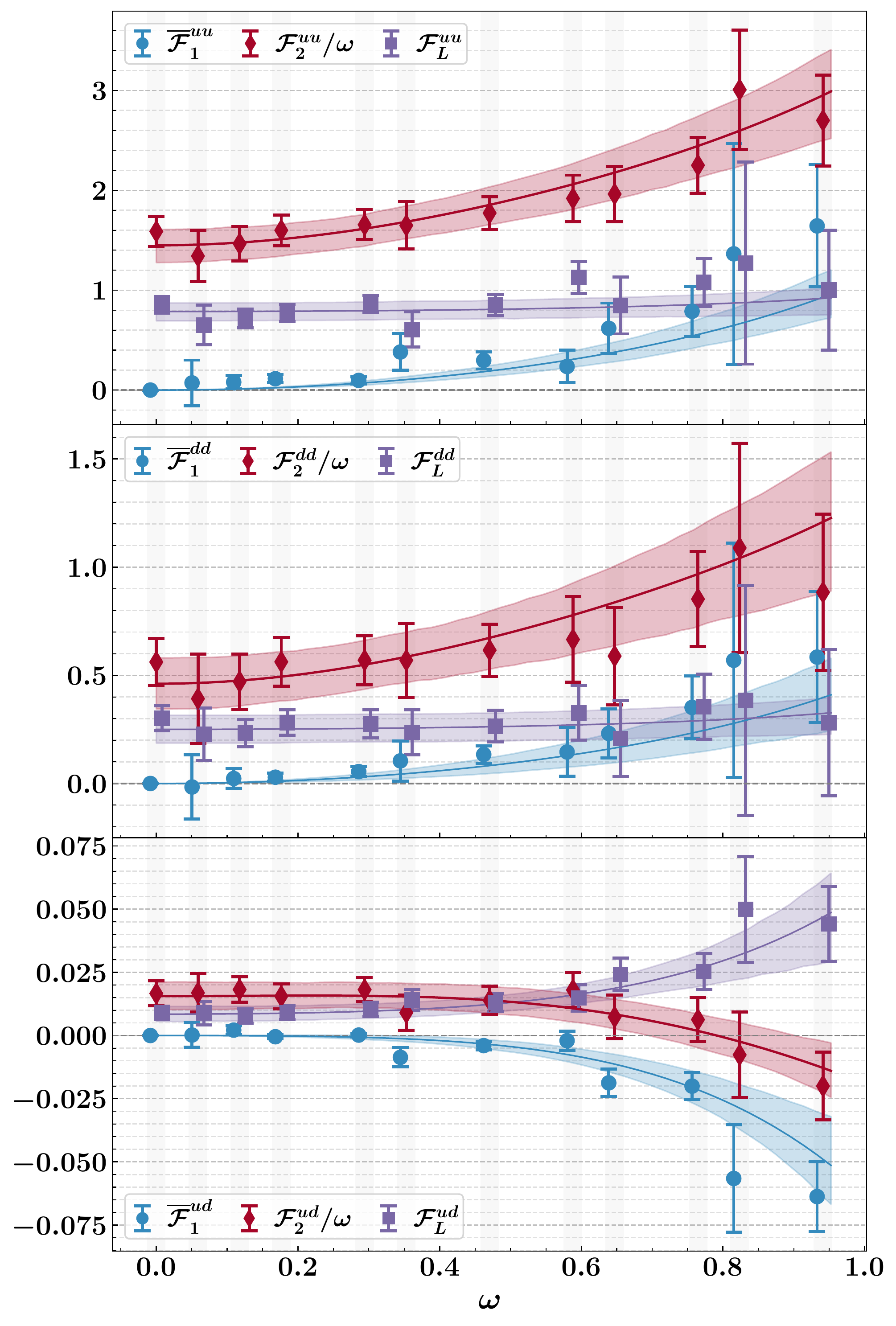}
    \caption{\label{fig:F12L}$\omega$ dependence of the Compton structure functions $\overline{\mathcal{F}}_1$, $\mathcal{F}_2$, and $\overline{\mathcal{F}}_L$ at $Q^2 = 4.86 \, {\rm GeV}^2$. We show the $uu$ (top), $dd$ (middle) and $ud$ (bottom) contributions. Coloured shaded bands show the fits with their 68\% credible region of the highest posterior density. Points are displaced for clarity. 
    }
\end{figure}

We show the $\omega$ dependence of the Compton structure functions along with their fit curves in \Cref{fig:F12L} for a representative case of $Q^2 = 4.86 \, {\rm GeV}^2$ calculated on the $48^3 \times 96$ ensemble. Note that a small (large) nucleon momentum $\bv{p}$ does not necessarily correspond to a small (large) $\omega$. This explains the larger uncertainties of some $\omega$ values (e.g. $\omega = 0.06, 0.35$ in \Cref{fig:F12L}) in comparison to their neighbours (see \Cref{fig:effmass_lamfit_2,fig:effmass_lamfit_3} in \Cref{app:enshfit} for a comparison of $\omega=0.06$ to $\omega=0.18$).

\begin{figure}[ht]
	\centering
	\includegraphics[width=.49\textwidth]{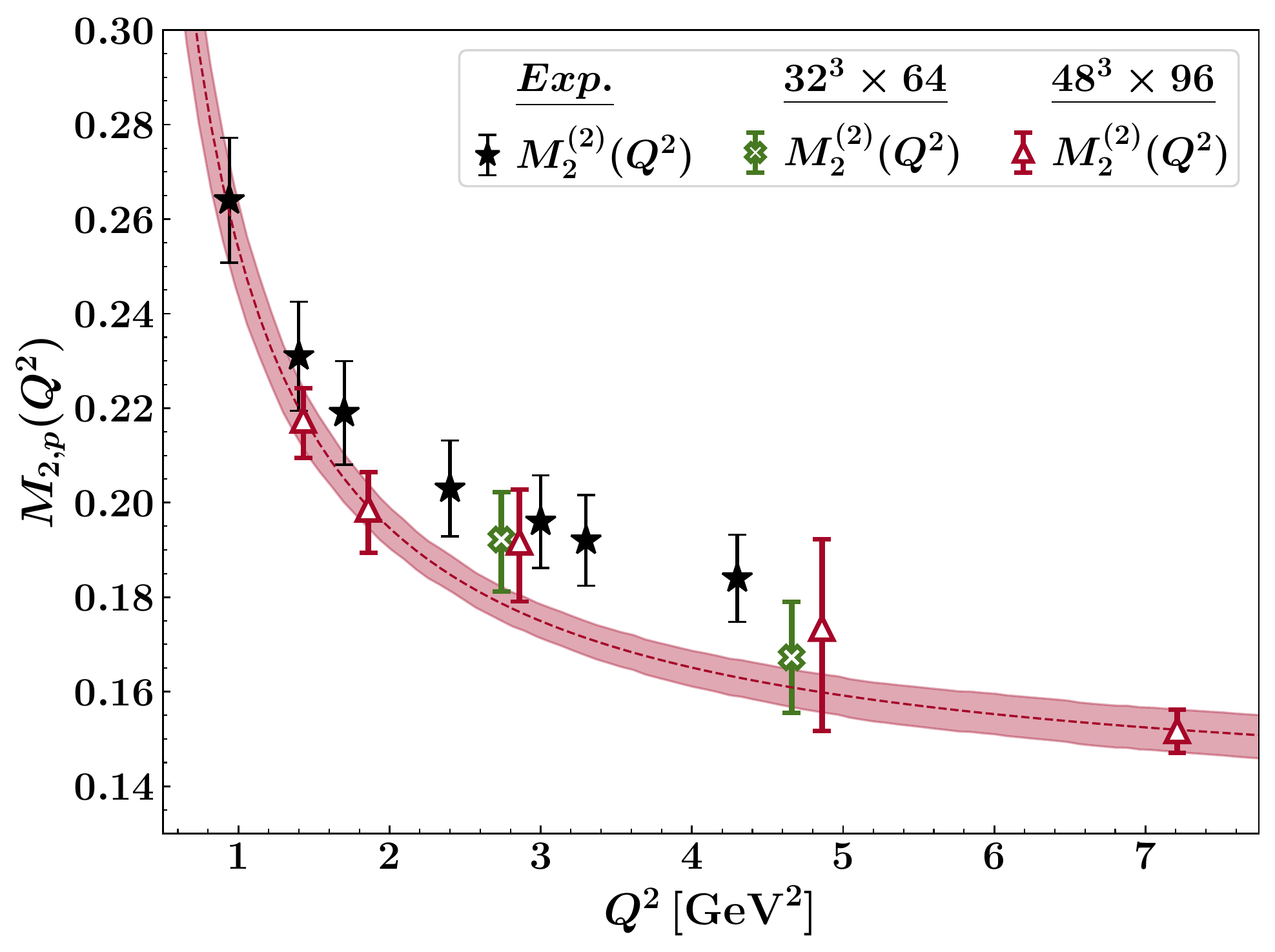}
	\caption{\label{fig:F2_proton_moments}$Q^2$ dependence of the lowest moments of $F_2$ for the proton. Filled stars are the experimental Cornwall-Norton moments of $F_2$ taken from Table I of Ref.~\cite{Armstrong:2001xj}. We have assigned a $5\%$ error to the experimental moments as indicated in Ref~\cite{Armstrong:2001xj}. Red band is the fit (\Cref{eq:ht_model}) to the $48^3\times96$ data points.}
\end{figure}

\begin{figure}[ht]
	\centering
	\includegraphics[width=.49\textwidth]{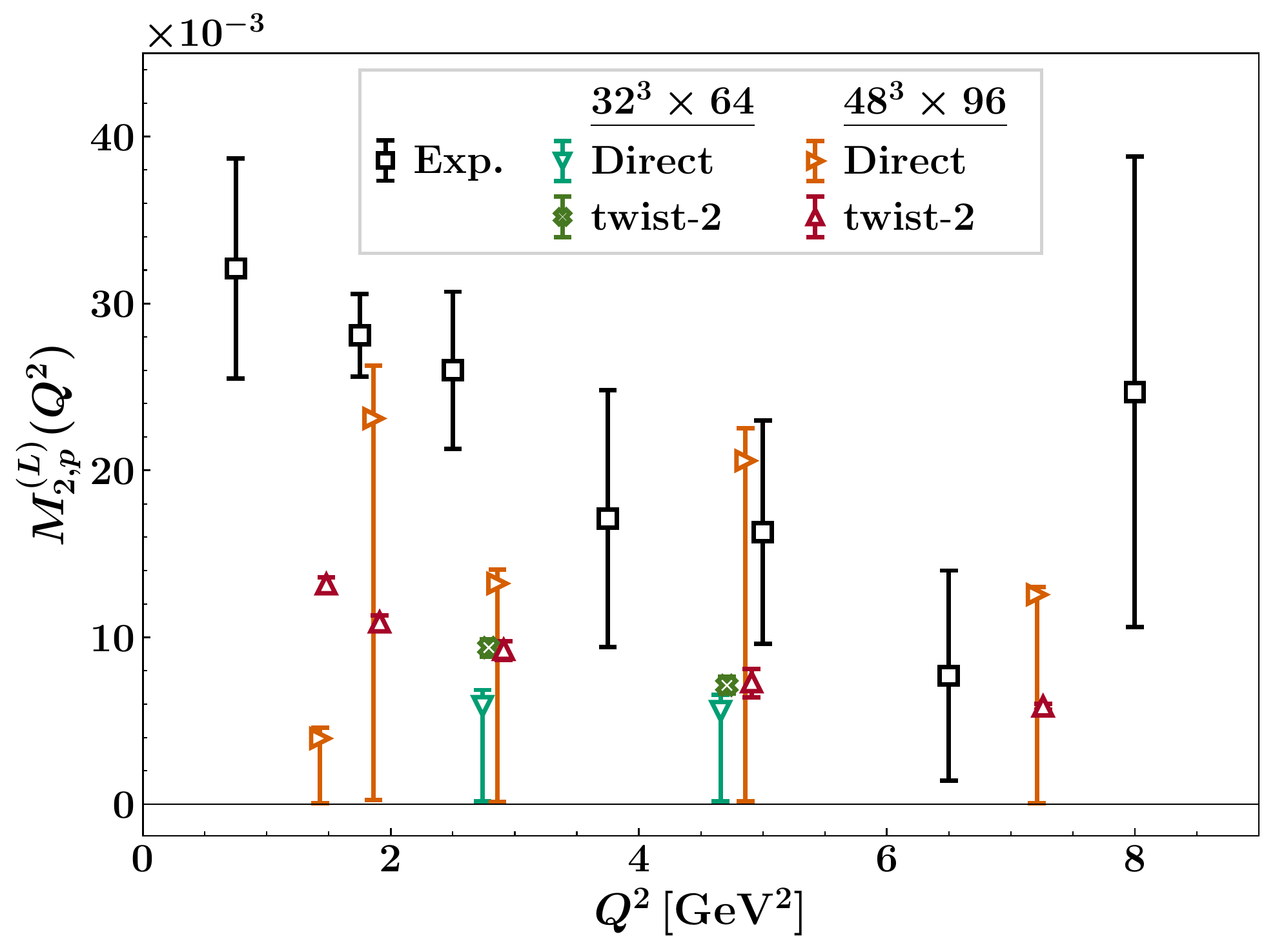}
	\caption{\label{fig:FL_proton_moments}Lowest moment of the proton's longitudinal structure function $M_{2,p}^{(L)}$ as a function of $Q^2$. We compare our results (Direct) to the experimental Nachtmann moments (open black squares) taken from~\cite{Monaghan:2012et}. Asymmetric error bars indicate that our posterior distributions are highly skewed (non-Gaussian). We also show the moments (twist-2) determined via the relation, \Cref{eq:FL_qcd}, using our determination of $M_{2,p}^{(2)}$ from the current work. Twist-2 points are displaced for clarity.
	}
\end{figure}

We keep terms up to $\mathcal{O}(w^8)$ in the fit polynomials~\Cref{eq:ope_moments1,eq:moments_F2}. The lowest two moments are insensitive to the addition of higher order terms (see \Cref{app:bayes}). 
The lowest moments of the structure functions $F_{2}$ and $F_{L}$ obtained from the $32^3\times64$ and $48^3\times96$ ensembles are shown in \Cref{fig:F2_proton_moments,fig:FL_proton_moments} as a function of $Q^2$ for the proton. Note that the moments of the proton are constructed via $M_{2,p}^{(2,L)} = \frac{4}{9} M_{2,uu}^{(2,L)} + \frac{1}{9} M_{2,dd}^{(2,L)} - \frac{2}{9} M_{2,ud}^{(2,L)}$. Our $F_{2}$ moments are in good agreement with the experimental moments~\cite{Armstrong:2001xj}, however, we remind the reader that our results do not yet incorporate chiral, infinite volume and continuum extrapolations. 

Since the Compton amplitude includes all power corrections, we can estimate the leading power correction (i.e. twist-4) by studying the $Q^2$ behaviour of the moments. Higher-twist contributions are suppressed by powers of $1/Q^2$ so one expects to have sizeable contributions for intermediate to low $Q^2$. Their effect (at the lowest order) can be modelled by the twist expansion,
\begin{equation}\label{eq:ht_model}
	M_{2,h}^{(2)}(Q^2) = M_{2,h}^{(2)} + C_{2,h}^{(2)}/Q^2 + \mathcal{O}(1/Q^4),	
\end{equation}
where $h \in \{uu, dd, ud, p\}$. We utilise only the $M_2^{(2)}(Q^2)$ moments obtained on the $48^3 \times 96$ ensemble down to $Q^2 \approx 1.5 \; {\rm GeV}^2$ to study the power corrections. We show our fit (\Cref{eq:ht_model}) in \Cref{fig:F2_proton_moments}. The extracted values for $M_{2,h}^{(2)}$ and $C_{2,h}^{(2)}$ are collected in \Cref{tab:model}. We note that our results could be useful for studies investigating the power corrections in the language of infrared renormalons~\cite{Stein1996,Dasgupta1996,Beneke2000}. 
\begin{table}[ht]
\centering
\caption{\label{tab:model} Extracted asymptotic values of the moments and the coefficients of the power correction terms. The power corrections are quoted at the scale of the nucleon mass $Q^2 = M_N^2$.}
\setlength{\extrarowheight}{2pt}
	\begin{tabular}{lcc}
		\hline\hline
		$h$ & $M_{2,h}^{(2)}$ & $C_{2,h}^{(2)} / M_N^2$ \\
		\hline
		$uu$ & 0.268(13) & 0.206(24) \\ 
		$dd$ & 0.146(7)  & 0.024(14) \\ 
		$ud$ & 0.000(0)  & 0.007(3)  \\ 
		$p$  & 0.135(6)  & 0.091(11) \\ 
		\hline\hline
	\end{tabular}
\end{table} 

We compare the lowest (Cornwall-Norton) moment of $F_L$ to the experimentally determined Nachtmann moments~\cite{Monaghan:2012et} in \Cref{fig:FL_proton_moments}. 
While we are unable to resolve a definitive signal for the $F_L$ moments, we are able to set an upper bound that is compatible with the experimental moments. 

It is interesting to compare $M_2^{(L)}$ determined from the relation~\cite{ALTARELLI197889},
\begin{equation}\label{eq:FL_qcd}
 	M_{2,p}^{(L),{\rm twist-}2}(Q^2) = \frac{4}{9\pi} \alpha_s(Q^2) M_{2,p}^{(2),{\rm twist-}2}(Q^2),
\end{equation}
where we replace the leading-twist moment on the RHS with $M_{2,p}^{(2)}(Q^2)$ from the current work as an approximation. We determine $\alpha_s(Q^2)$ at the four-loop order by running its value from the reference $M_\tau$ (tau-mass) scale that is extracted directly from $\tau$ decays~\cite{Baikov:2008jh} with $n_f = 3$ active flavours. The \texttt{CRunDec} package~\cite{Chetyrkin:2000yt,Herren:2017osy} is used to run the strong coupling constant. The effects of the number of active flavours, running from the $M_Z$ scale as opposed to $M_\tau$ scale, and crossing the charm quark threshold are negligible at this stage in contrast to the large uncertainties of experimental and lattice data.

The $Q^2$ behaviour is in good agreement with experimental points as shown in \Cref{fig:FL_proton_moments}. With improved precision in future studies, contrasting the direct determination and twist-2 part of the lowest few moments of $F_L$ would provide improved constraints on higher-twist effects.

\section{Conclusions} 
We have presented results of the lowest moments of the proton structure functions $F_2$ and $F_L$ as a function of $Q^2$, ranging from $Q^2 \approx 1\,\textrm{GeV}^2$ to $Q^2 \approx 7\,\textrm{GeV}^2$. The calculations have been done at the SU(3) flavour symmetrical point. This has been possible for the first time on the lattice, due to recent advances in computing the forward Compton amplitude using the second-order Feynman-Hellmann theorem. Power corrections turn out to be significant, up to $Q^2 \approx 5\,\textrm{GeV}^2$, and much larger than anticipated in theoretical estimates~\cite{Alekhin:2012ig,Dasgupta1996}. Already at unphysical quark masses we find good agreement with the moments extracted from experiment. However, calculations on additional ensembles that cover a range of lattice spacings and pion masses are required to fully account for systematic effects and rigorously confirm our findings. Our results are encouraging and show the potential of this approach to nucleon structure, starting from the all-encompassing Compton amplitude. The next natural step is to quantify the lattice systematics. Beyond the unpolarised structure, we are working towards extending our formalism to include the spin-dependent structure functions. Additionally, applying this method to the parity violating sector by considering weak currents is an exciting future direction.

\acknowledgments
We would like to thank Wally Melnitchouk for fruitful discussions. The numerical configuration generation (using the BQCD lattice QCD program~\cite{Haar:2017ubh})) and data analysis (using the Chroma software library~\cite{Edwards:2004sx}) was carried out on the DiRAC Blue Gene Q and Extreme Scaling (EPCC, Edinburgh, UK) and Data Intensive (Cambridge, UK) services, the GCS supercomputers JUQUEEN and JUWELS (NIC, Jülich, Germany) and resources provided by HLRN (The North-German Supercomputer Alliance), the NCI National Facility in Canberra, Australia (supported by the Australian Commonwealth Government) and the Phoenix HPC service (University of Adelaide). RH is supported by STFC through grant ST/P000630/1. PELR is supported in part by the STFC under contract ST/G00062X/1. KUC, RDY and JMZ are supported by the Australian Research Council grants DP190100297 and DP220103098. For the purpose of open access, the authors have applied a Creative Commons Attribution (CC BY) licence to any Author Accepted Manuscript version arising from this submission.

\appendix
\section{Extracting the energy shifts} \label{app:enshfit}
We form the ratios defined in \Cref{eq:ratio_fm,eq:ratio_fd} in order to extract the energy shifts from the perturbed correlators. The allowed $\bv{p}$ momenta are limited by $\bv{p}^2 \le [5, 5, 5, 10, 17]$ in lattice units for $\bv{q} = [(3,1,0)$, $(3,2,0)$, $(4,2,0)$, $(5,3,0)$, $(7,1,0)] \, \latmom$, respectively. Higher $\bv{p}^2$ cuts introduce duplicates of $\omega$ values with worsening signal quality, thus do not expand the $\omega$ coverage any further. We tabulate the used $\omega$ values in \Cref{tab:pmom}. We omit some high-$\bv{p}$ momenta in the analysis due to their poor S/N which hinders a reliable extraction of the ground state energy shifts. It is possible to improve the signal quality of such higher momenta correlators by employing momentum smearing techniques~\cite{PhysRevD.93.094515}, which we plan to investigate in future work.

\begin{table}[h]
\centering
\caption{\label{tab:pmom} Multiple $\omega$ values that we can access with several combinations of $\bv{p} = (p_1,p_2,p_3)$ and $\bv{q} = (q_1,q_2,q_3)$ in lattice units, where we have set $p_3 = q_3 = 0$. We only show the $(\bv{p},\bv{q})$ combinations that give a positive $\omega$. The $\omega \ge \emph{1}$ values (indicated by italics) are omitted since they lie outside the allowed $\omega$ range. The regular typeset $\omega$ values are also omitted due to their poor signal quality. We use the $\omega$ values shown in boldface only.}
\setlength{\extrarowheight}{2pt}
\begin{tabularx}{.485\textwidth}{L|CCCCC}
	\hline\hline
	& \multicolumn{5}{c}{$\omega = 2 \bv{p}\cdot\bv{q}/Q^2$} \\
	\hline
	\multirow{2}{*}{$\bv{p} / (2\pi/L)$}  & \multicolumn{5}{c}{$\bv{q} / (2\pi/L)$} \\
	& $(3,1,0)$ & $(3,2,0)$ & $(4,2,0)$ & $(5,3,0)$ & $(7,1,0)$ \\
	\hline
	(0, 0, 0)  &  {\bf 0.0} &  {\bf 0.00} 	&  {\bf 0.0} &  {\bf 0.00}  &  {\bf 0.00} \\
	(0, 1, 0)  &  {\bf 0.2} &  {\bf 0.31} 	&  {\bf 0.2} &  {\bf 0.18}  &  {\bf 0.04} \\
	(0, 2, 0)  &  0.4 		&  {\bf 0.62} 	&  0.4 		 &  {\bf 0.35}  &  {\bf 0.08} \\
	(0, 3, 0)  &  --- 		&   --- 		&  --- 		 &  0.53 		&  {\bf 0.12} \\
	(0, 4, 0)  &  --- 		&   --- 		&  --- 		 &   --- 		&  {\bf 0.16} \\
	(1, 0, 0)  &  {\bf 0.6} &  {\bf 0.46} 	&  {\bf 0.4} &  {\bf 0.29}  &  {\bf 0.28} \\
	(1, 1, 0)  &  {\bf 0.8} &  {\bf 0.77} 	&  {\bf 0.6} &  {\bf 0.47}  &  {\bf 0.32} \\
	(1, 2, 0)  &  {\it 1.0} &  {\it 1.08} 	&  {\bf 0.8} &  {\bf 0.65}  &  {\bf 0.36} \\
	(1, 3, 0)  &  --- 		&   --- 		&  --- 		 &  {\bf 0.82}  &  {\bf 0.40} \\
	(1, 4, 0)  &  --- 		&   --- 		&  --- 		 &   --- 		&  0.44 \\
	(1, -1, 0) &  {\bf 0.4} &  {\bf 0.15} 	&  {\bf 0.2} &  {\bf 0.12} 	&  {\bf 0.24} \\
	(1, -2, 0) &  0.2 		&   --- 		&  0.0 		 &   --- 		&  {\bf 0.20} \\
	(-1, 2, 0) &  --- 		&  {\bf 0.15} 	&  0.0 		 &  {\bf 0.06} 	&   --- \\
	(1, -3, 0) &  --- 		&   --- 		&  --- 		 &   --- 		&  {\bf 0.16} \\
	(-1, 3, 0) &  --- 		&   --- 		&  --- 		 &  0.24 		&   --- \\
	(1, -4, 0) &  --- 		&   --- 		&  --- 		 &   --- 		&  0.12 \\
	(2, 0, 0)  &  {\it 1.2} &  {\bf 0.92} 	&  {\bf 0.8} &  {\bf 0.59} 	&  {\bf 0.56} \\
	(2, 1, 0)  &  {\it 1.4} &  {\it 1.23} 	&  {\it 1.0} &  {\bf 0.76} 	&  {\bf 0.60} \\
	(2, 2, 0)  &  --- 		&   --- 		&  --- 		 &  {\bf 0.94} 	&  {\bf 0.64} \\
	(2, 3, 0)  &  --- 		&   --- 		&  --- 		 &   --- 		&  0.68 \\
	(2, -1, 0) &  {\it 1.0} &  {\bf 0.62} 	&  0.6 		 &  0.41 		&  {\bf 0.52} \\
	(2, -2, 0) &  --- 		&   --- 		&  --- 		 &  0.24 		&  {\bf 0.48} \\
	(2, -3, 0) &  --- 		&   --- 		&  --- 		 &   --- 		&  0.44 \\
	(3, 0, 0)  &  --- 		&   --- 		&  --- 		 &  0.88 		&  {\bf 0.84} \\
	(3, 1, 0)  &  --- 		&   --- 		&  --- 		 &  {\it 1.06} 	&  {\bf 0.88} \\
	(3, 2, 0)  &  --- 		&   --- 		&  --- 		 &   --- 		&  0.92 \\
	(3, 3, 0)  &  --- 		&   --- 		&  --- 		 &   --- 		&  0.96 \\
	(3, -1, 0) &  --- 		&   --- 		&  --- 		 &  0.71 		&  {\bf 0.80} \\
	(3, -2, 0) &  --- 		&   --- 		&  --- 		 &   --- 		&  0.76 \\
	(3, -3, 0) &  --- 		&   --- 		&  --- 		 &   --- 		&  0.72 \\
	(4, 0, 0)  &  --- 		&   --- 		&  --- 		 &   --- 		&  {\it 1.12} \\
	(4, 1, 0)  &  --- 		&   --- 		&  --- 		 &   --- 		&  {\it 1.16} \\
	(4, -1, 0) &  --- 		&   --- 		&  --- 		 &   --- 		&  {\it 1.08} \\
	\hline\hline
\end{tabularx}
\end{table}

\begin{figure*}[ht]
    \centering
    \includegraphics[width=.6\textwidth]{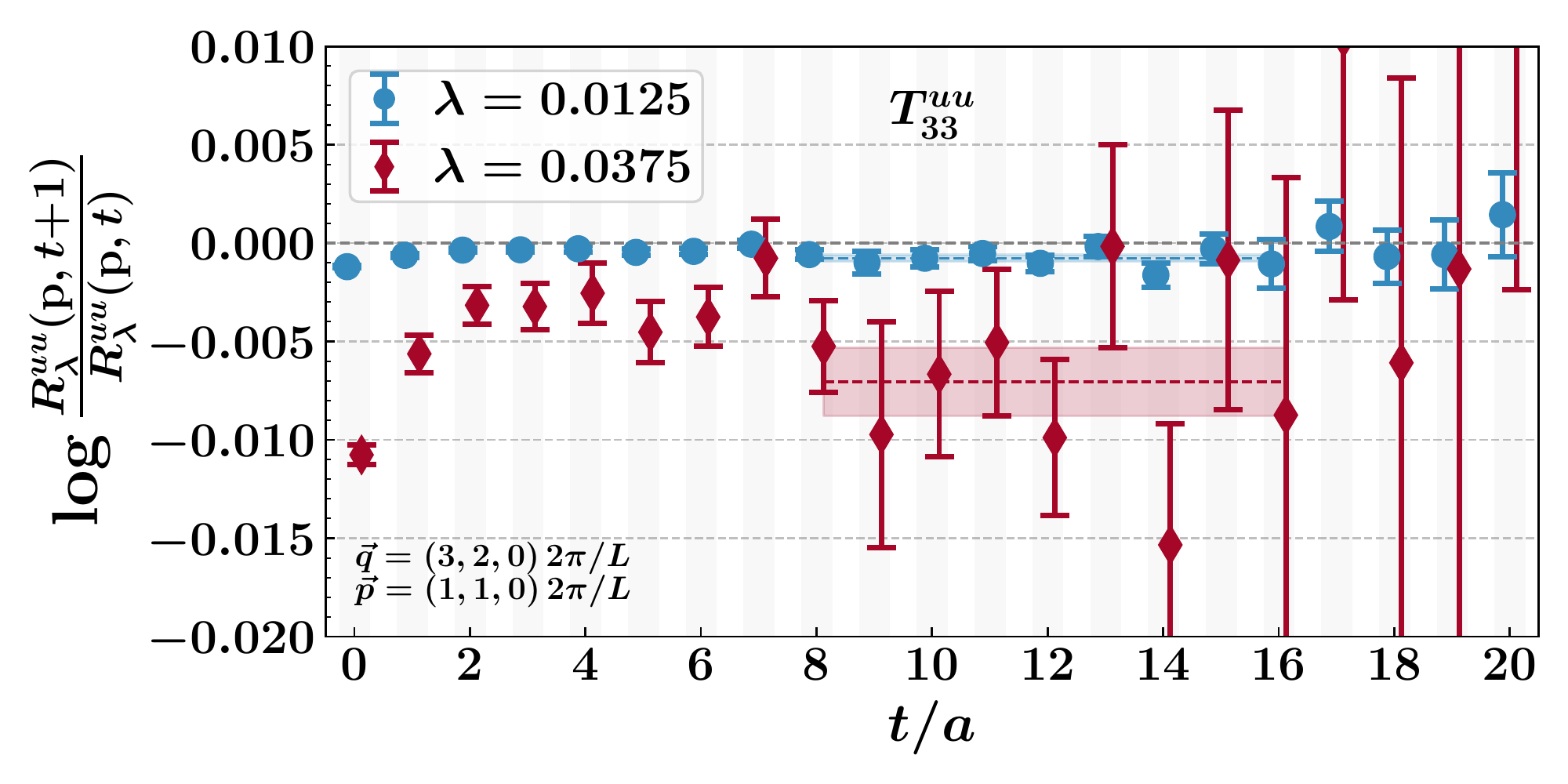} \\
    \includegraphics[width=.6\textwidth]{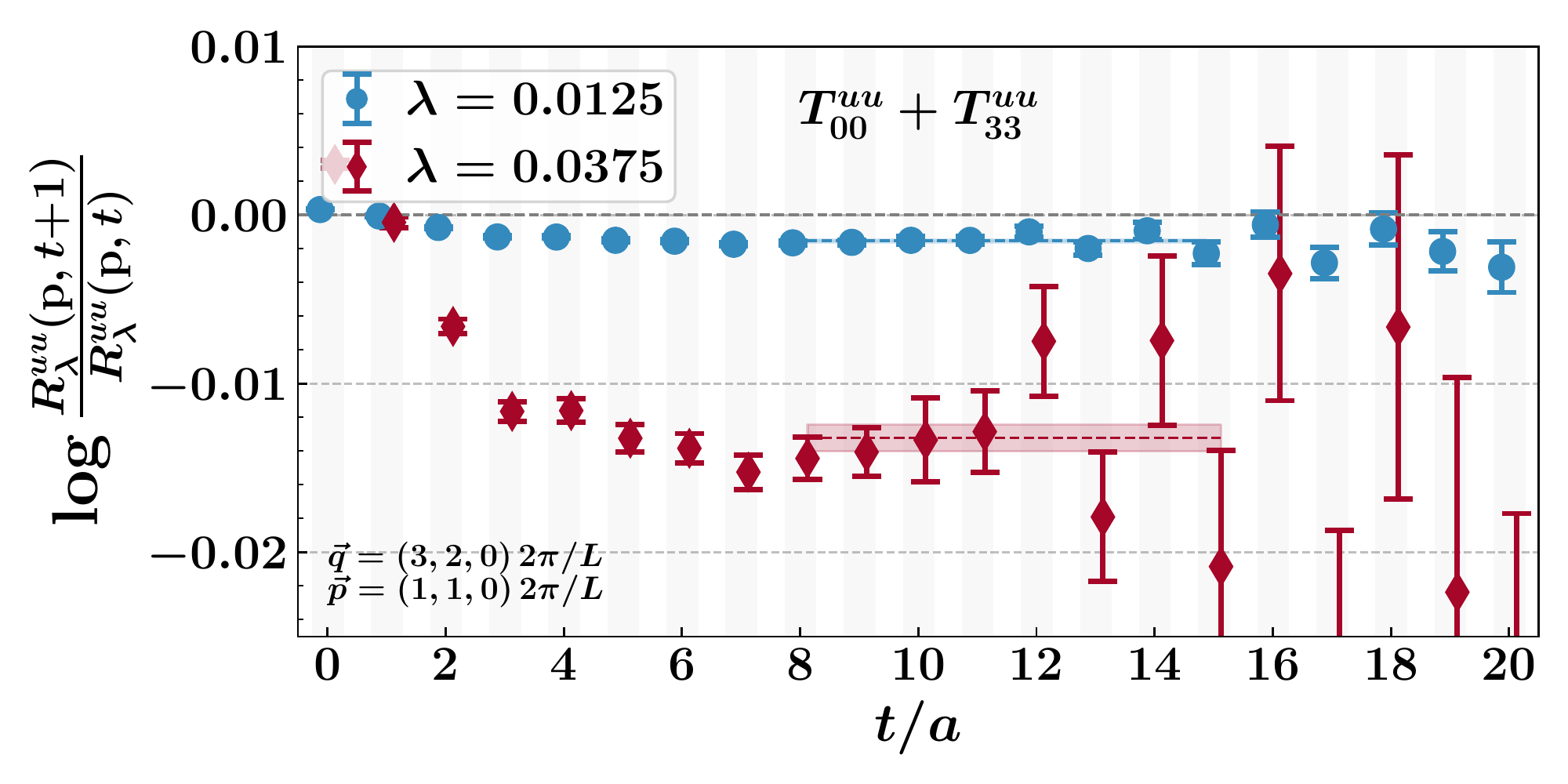} \\
    \includegraphics[width=.6\textwidth]{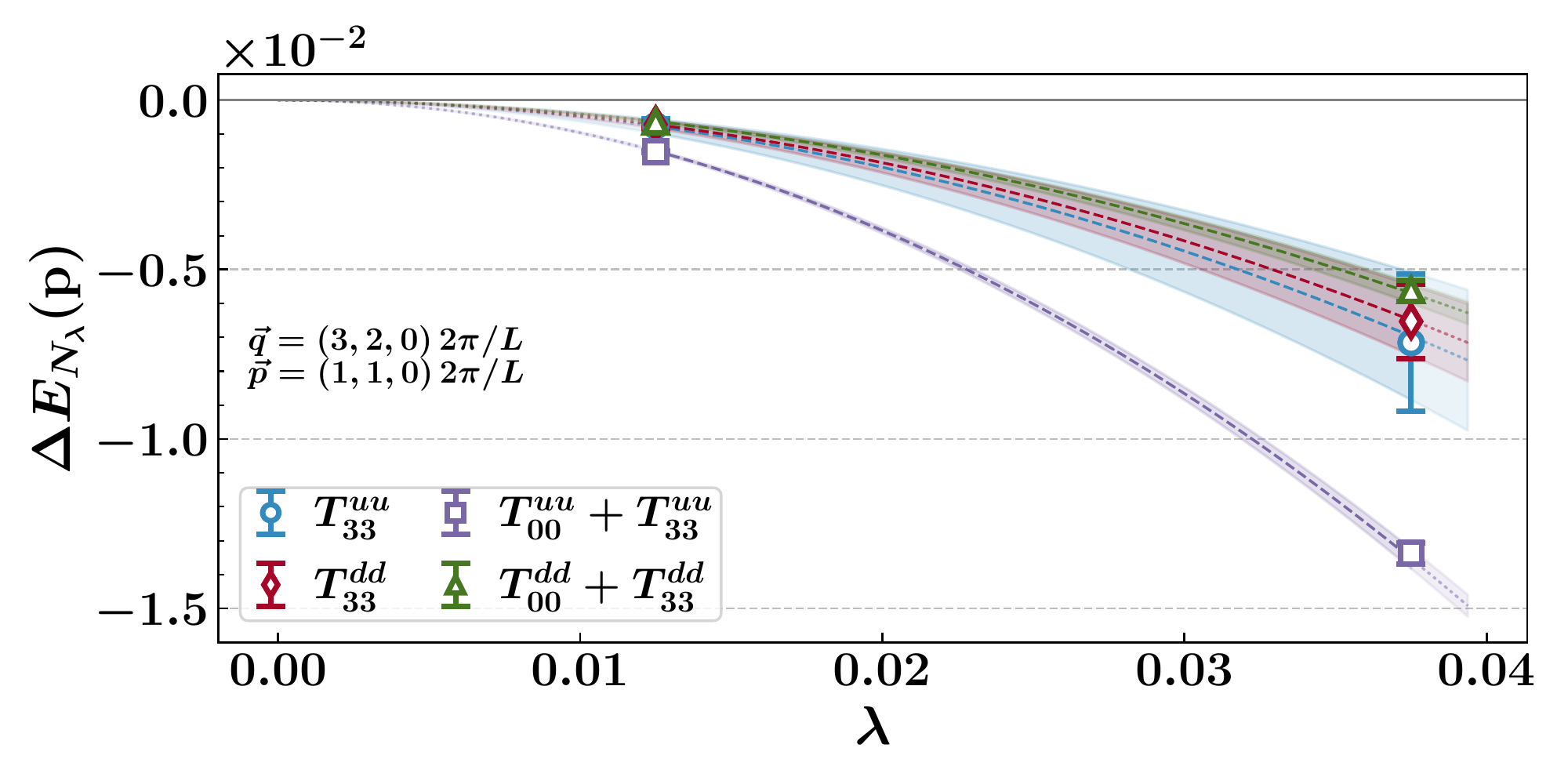}
    \caption{\label{fig:effmass_lamfit_1} Top to bottom: Effective mass plots for the correlator ratios of the amplitudes $T_{33}$ and $T_{00}+T_{33}$, and the corresponding fits in $\lambda$-space, respectively. Shaded regions on the correlator ratio plots depict the fit windows and extracted energy shifts with their $1\sigma$ uncertainty bands. Shaded curves on the $\lambda$-fit plots indicate the fit curves and their $1\sigma$ uncertainties. We show the results for $\omega=0.77$ ($\bv{p}=(1,1,0) \, \latmom$) for $\bv{q}=(3,2,0) \, \latmom$ obtained on the $48^3 \times 96$ ensemble.
    }
\end{figure*}

\begin{figure*}[ht]
	\includegraphics[width=.6\textwidth]{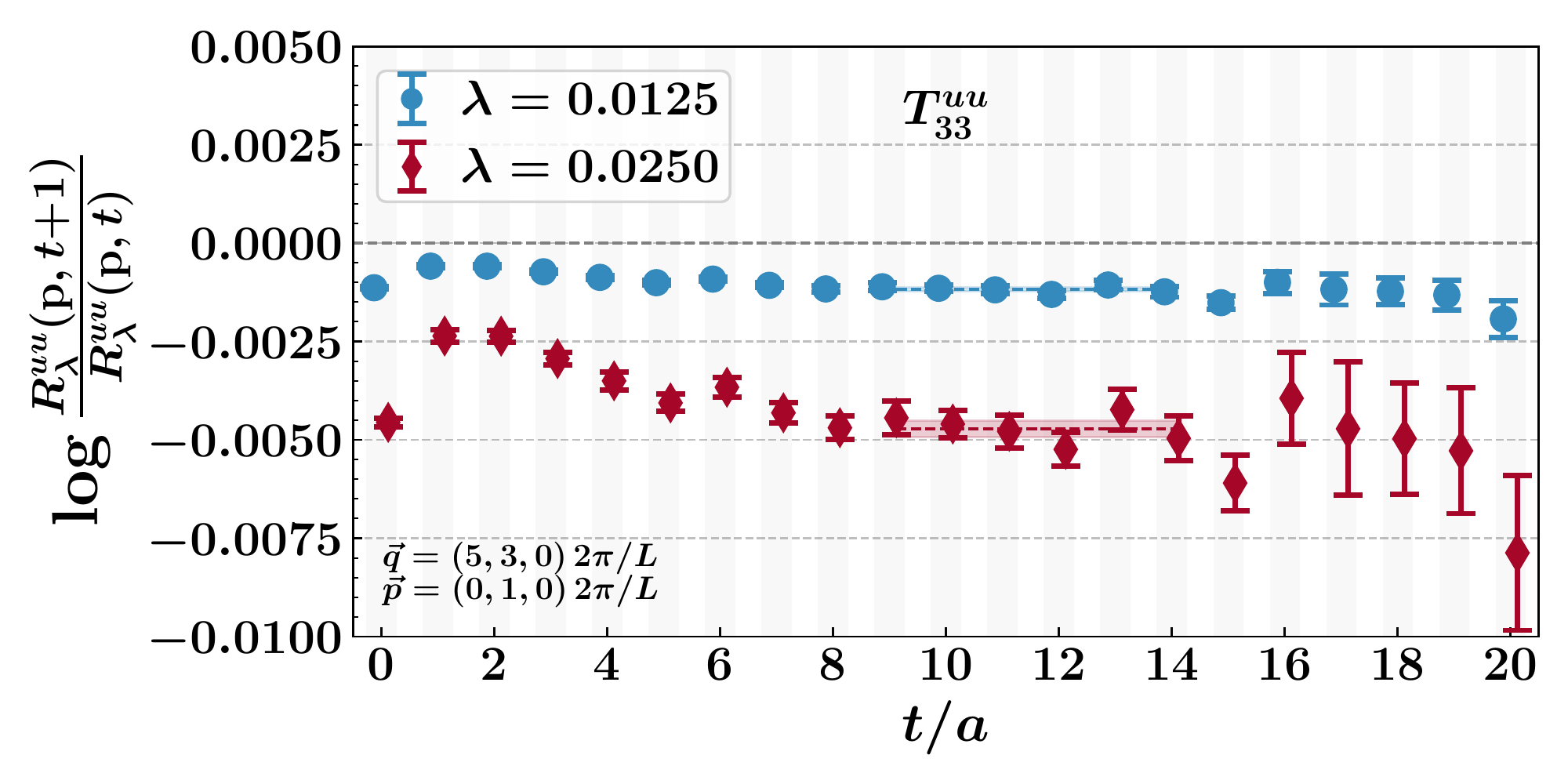} \\
	\includegraphics[width=.6\textwidth]{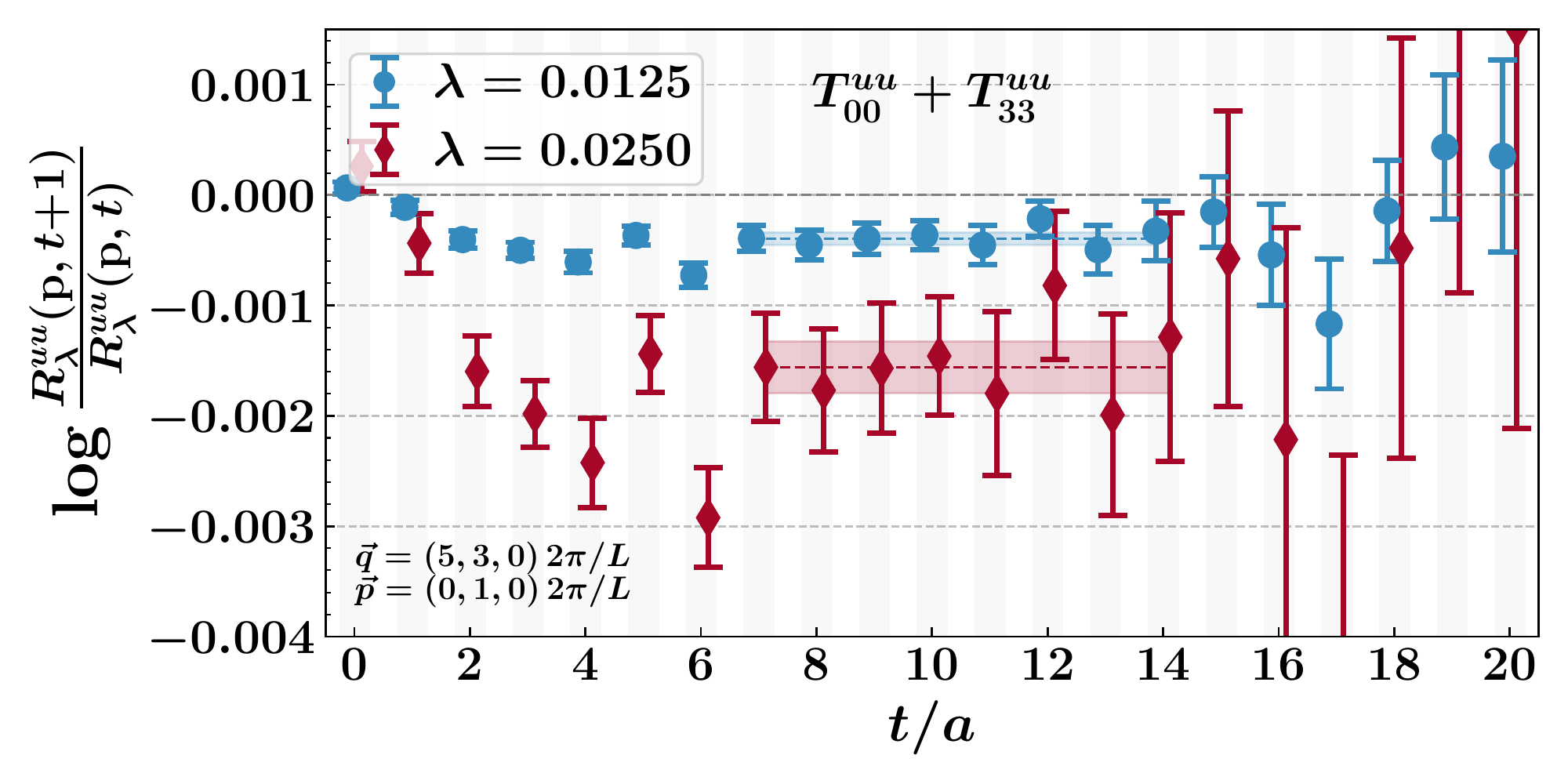} \\
    \includegraphics[width=.6\textwidth]{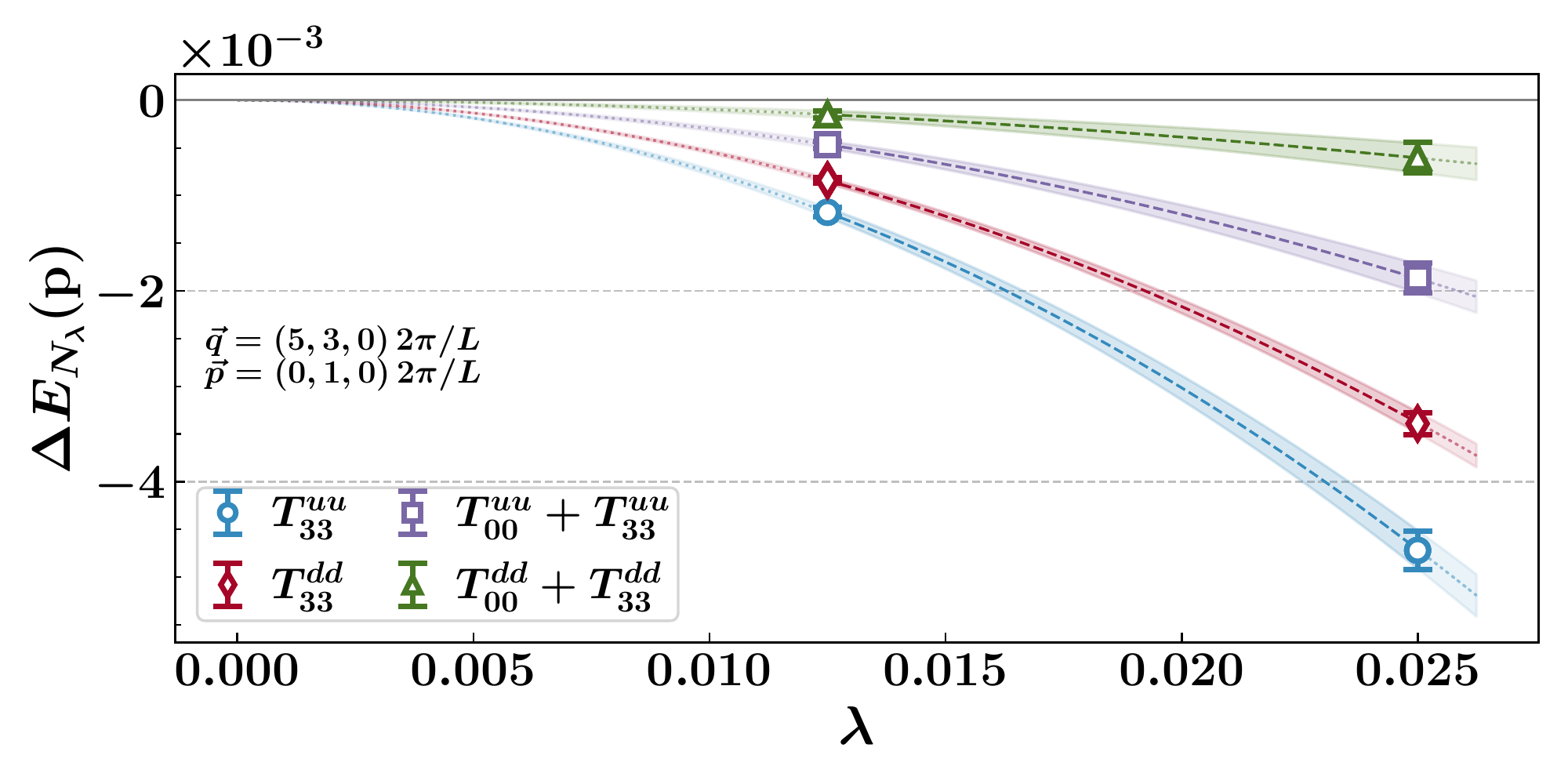}
    \caption{\label{fig:effmass_lamfit_2} Same as \Cref{fig:effmass_lamfit_1} but for $\omega=0.18$ ($\bv{p}=(0,1,0) \, \latmom$) for $\bv{q}=(5,3,0) \, \latmom$.
    }
\end{figure*}

\begin{figure*}[ht]
    \includegraphics[width=.6\textwidth]{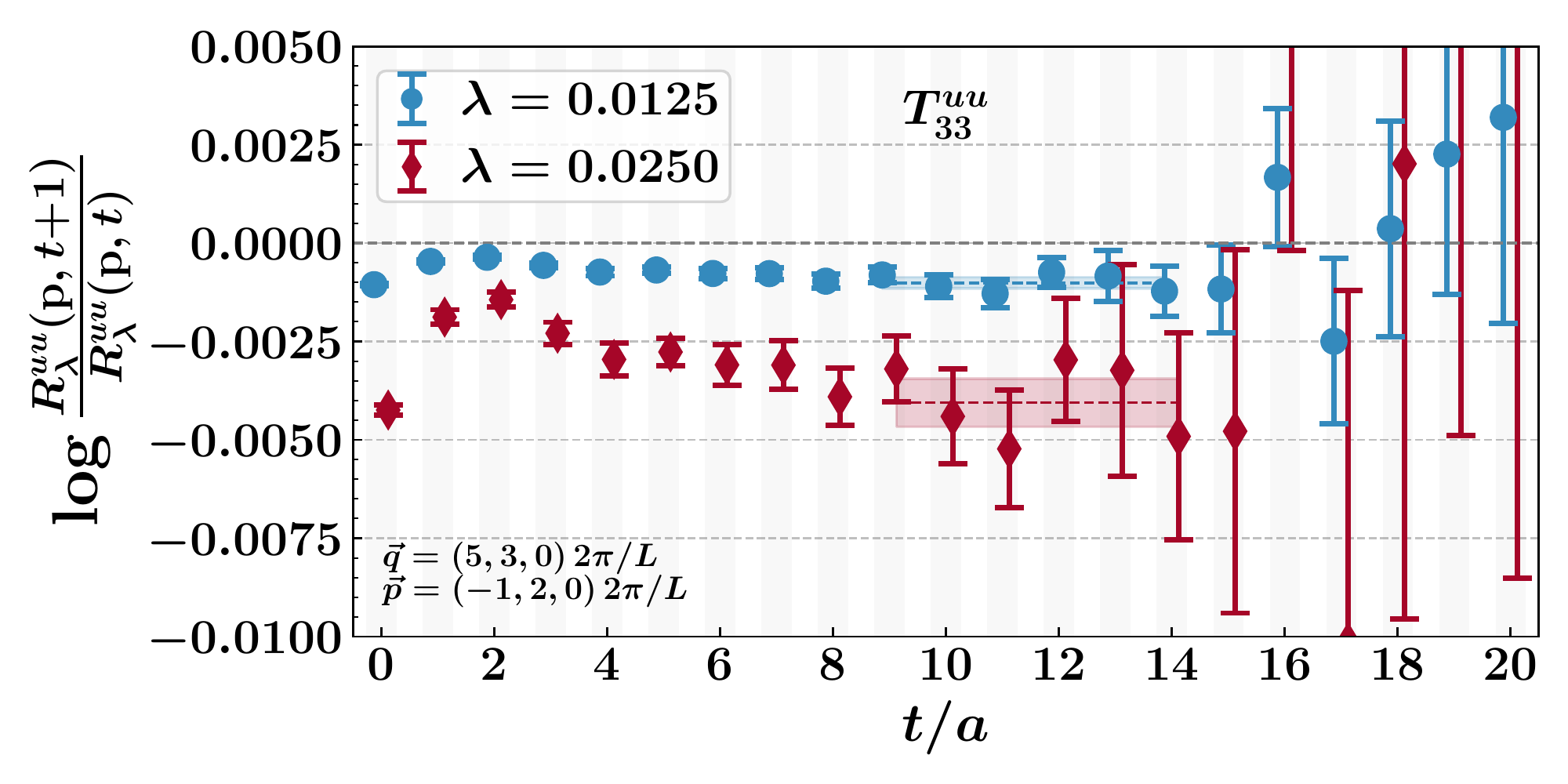} \\
    \includegraphics[width=.6\textwidth]{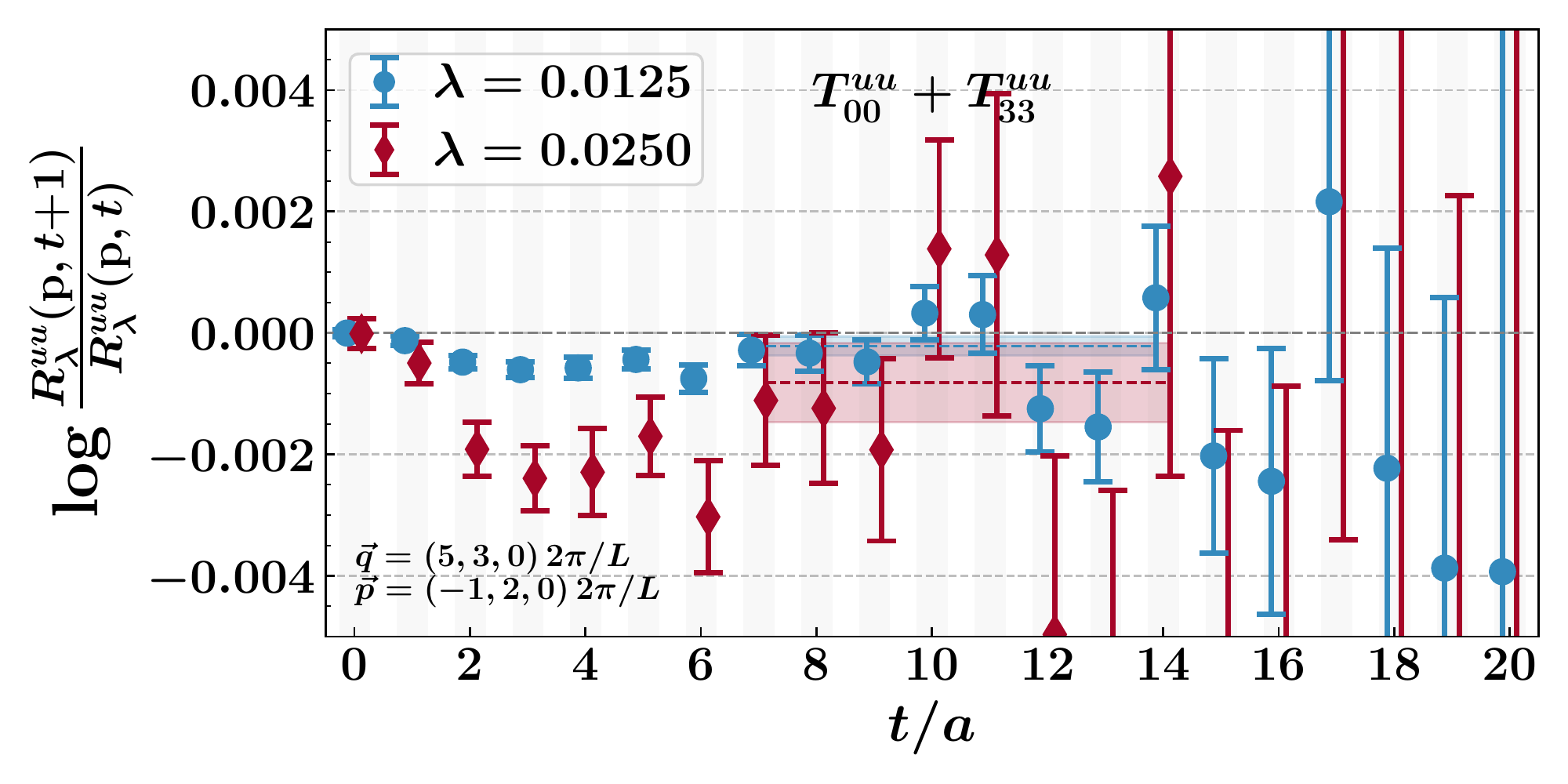} \\
    \includegraphics[width=.6\textwidth]{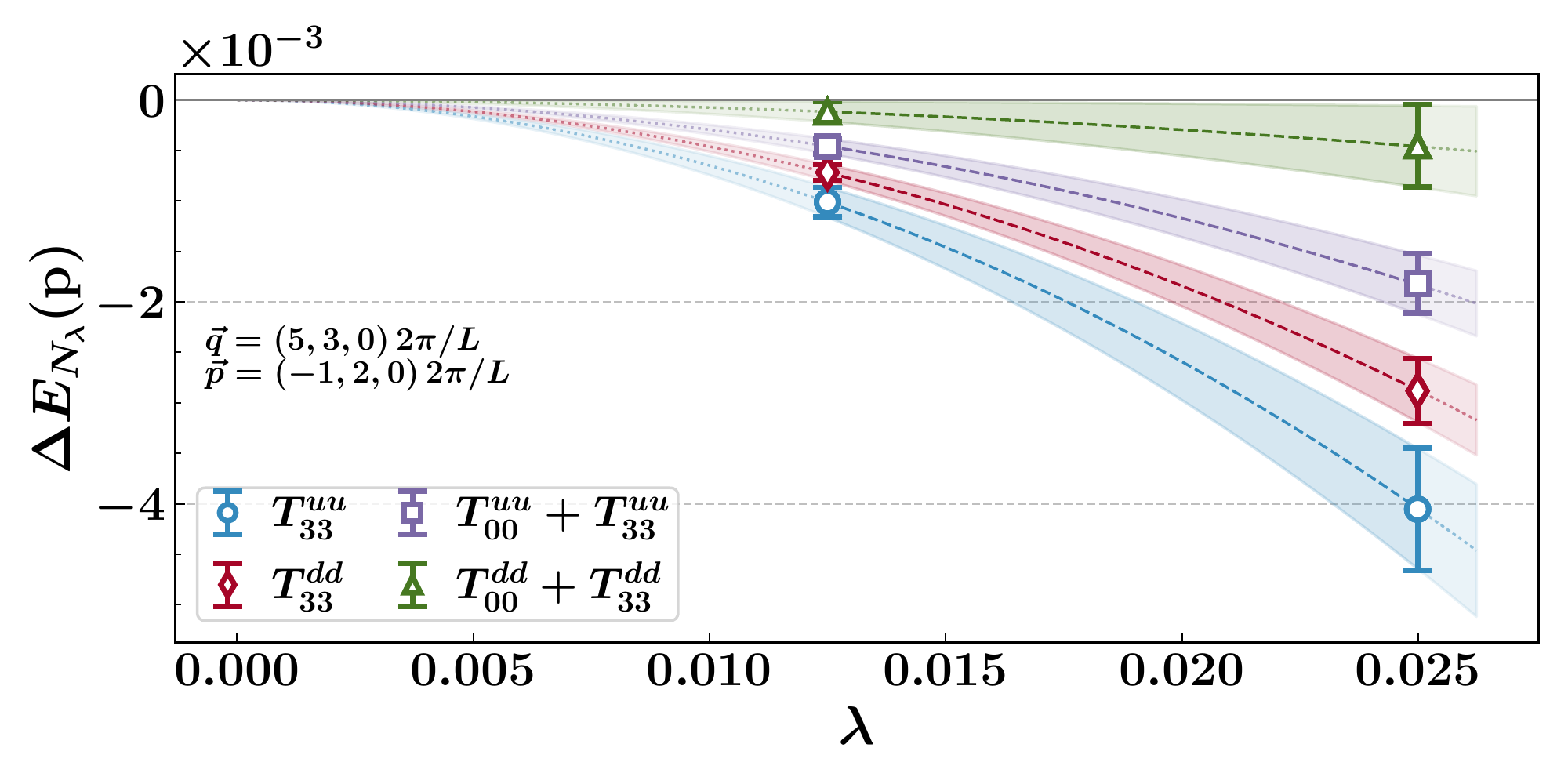}
    \caption{\label{fig:effmass_lamfit_3} Same as \Cref{fig:effmass_lamfit_1} but for $\omega=0.06$ ($\bv{p}=(-1,2,0) \, \latmom$) for $\bv{q}=(5,3,0) \, \latmom$.
    }
\end{figure*}

\begin{figure*}[ht]
	\centering
	\includegraphics[width=.6\textwidth]{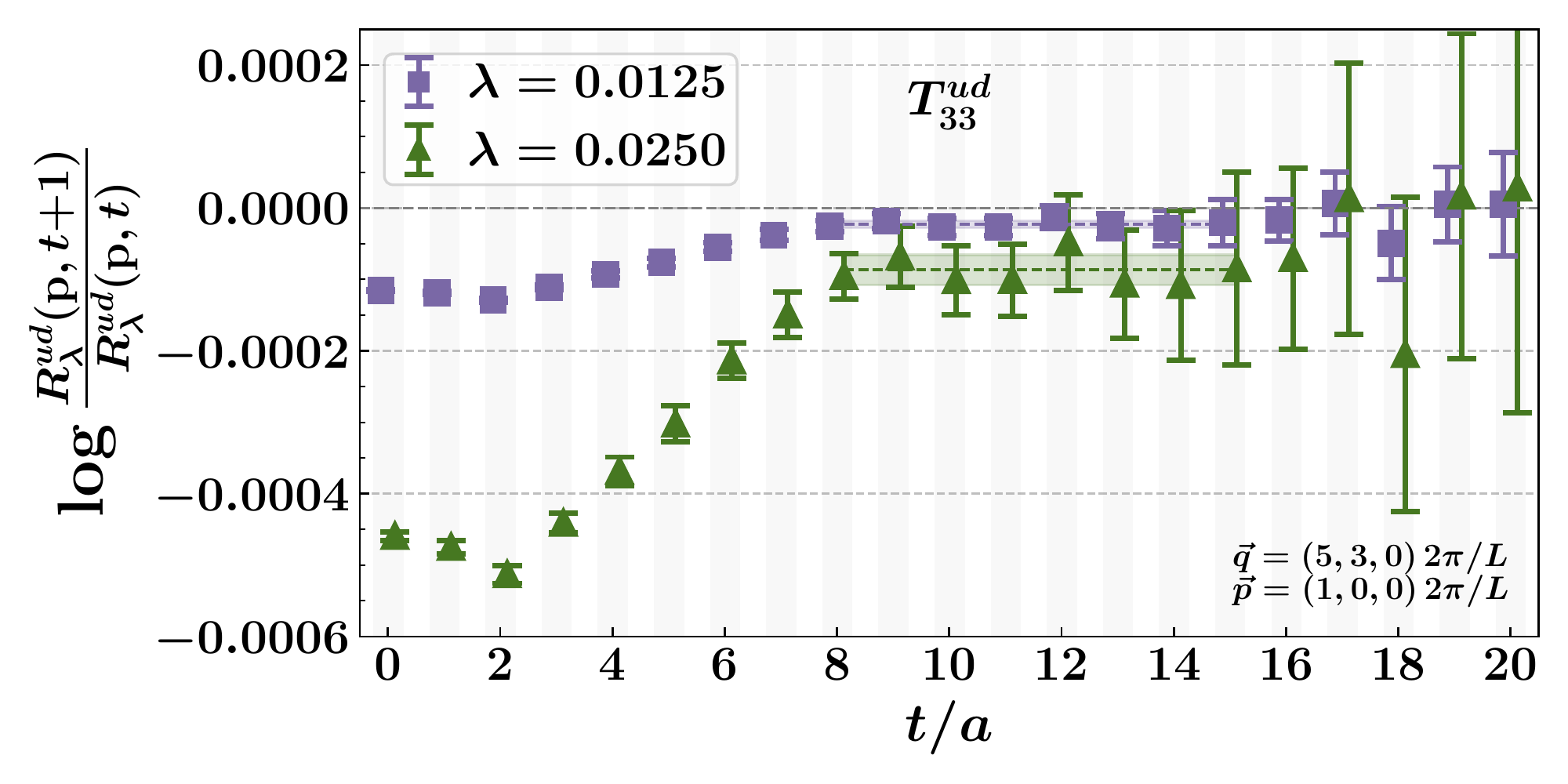} \\
	\includegraphics[width=.6\textwidth]{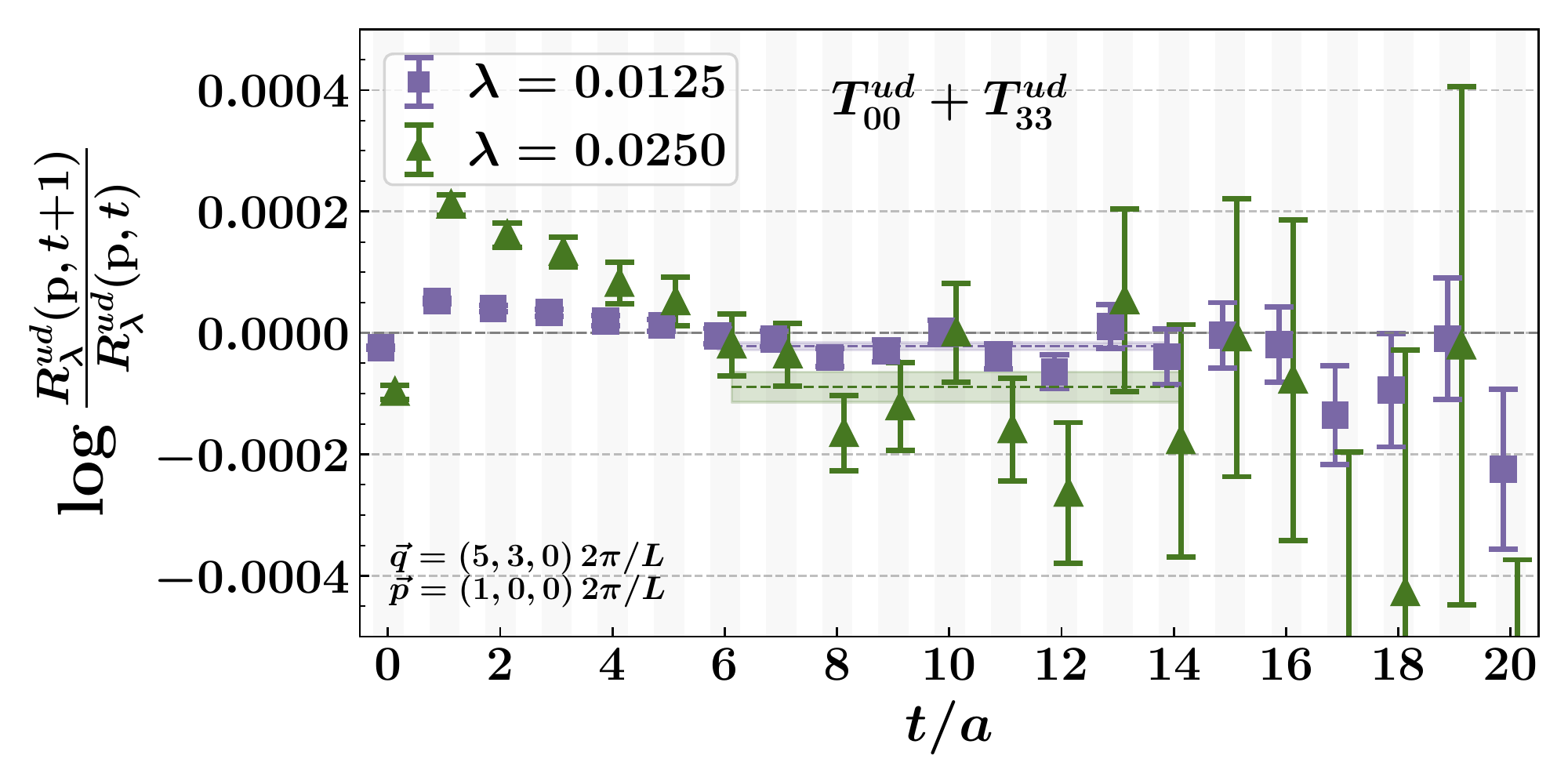} \\
	\includegraphics[width=.6\textwidth]{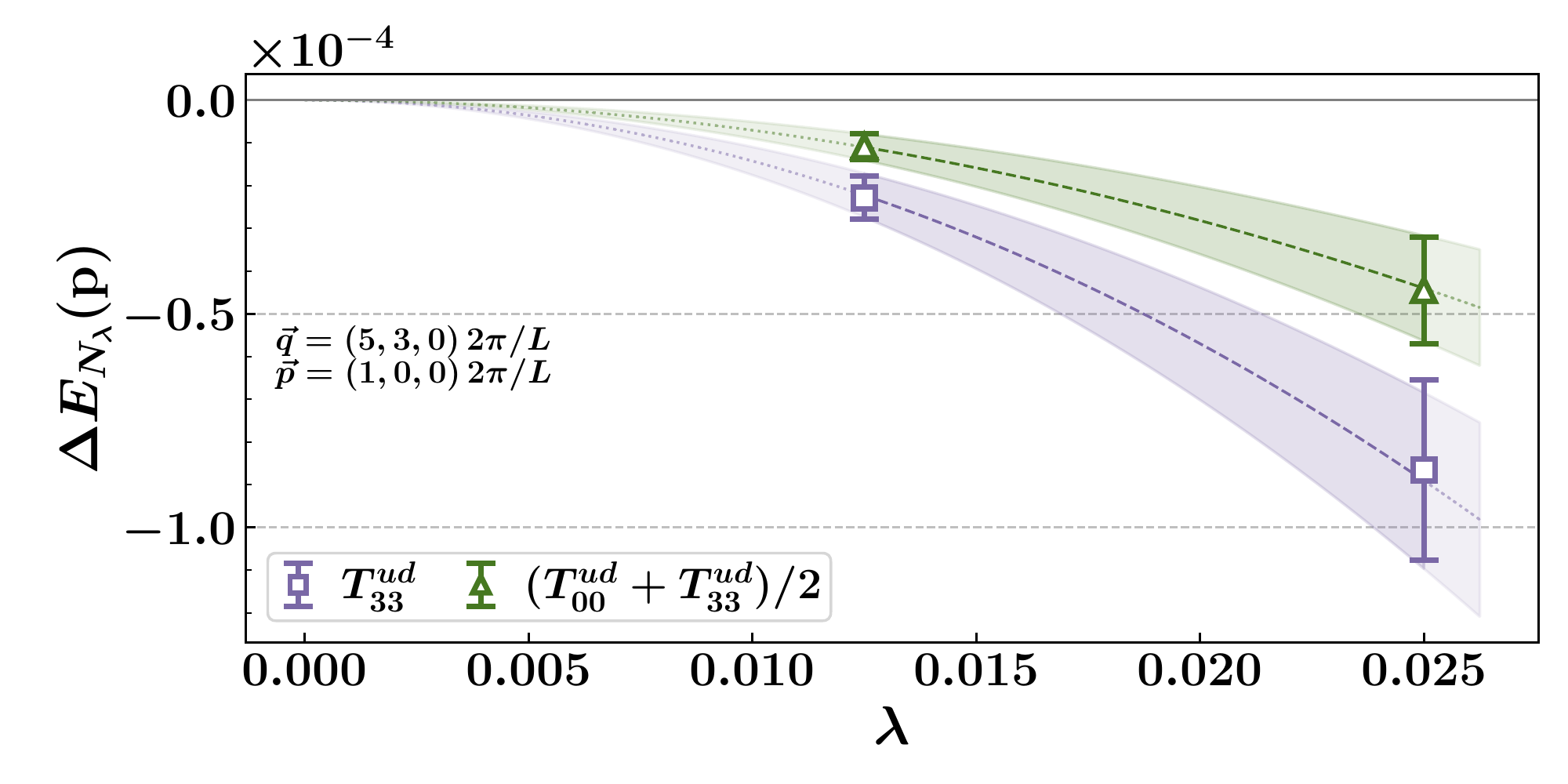}
	\caption{\label{fig:effmass_lamfit_ud} Same as \Cref{fig:effmass_lamfit_1} but for the purely higher-twist $ud$ piece. We show the results obtained on the $48^3 \times 96$ ensemble for the $(\bv{p},\bv{q}) = ((1,0,0),\, (5,3,0)) \, \latmom$ pair. Energy shifts for the $T_{00}+T_{33}$ combination have been rescaled by a factor of $0.5$ on the bottom plot for clarity.}
\end{figure*}

Effective mass plots for the correlator ratios are shown in \Cref{fig:effmass_lamfit_1,fig:effmass_lamfit_2,fig:effmass_lamfit_3} along with the fits performed in $\lambda$-space to extract the energy shifts for three different kinematics. We show the ratios for the $uu$ piece only. The $dd$ piece behaves similarly. Analogous plots for the $ud$ piece are shown in \Cref{fig:effmass_lamfit_ud}. The $\mathcal{F}_1$ amplitude is isolated from $T_{33}$ in a straightforward fashion, while the $\mathcal{F}_2/\omega$ amplitude is accessed from the $T_{00} + T_{33}$ combination up to known kinematical factors (\Cref{eq:compF2}). Top (middle) rows of \Cref{fig:effmass_lamfit_1,fig:effmass_lamfit_2,fig:effmass_lamfit_3} show the the correlator ratios for $T_{33}$ ($T_{00} + T_{33}$). 

\begin{figure*}[ht]
	\centering
	\includegraphics[width=.6\textwidth]{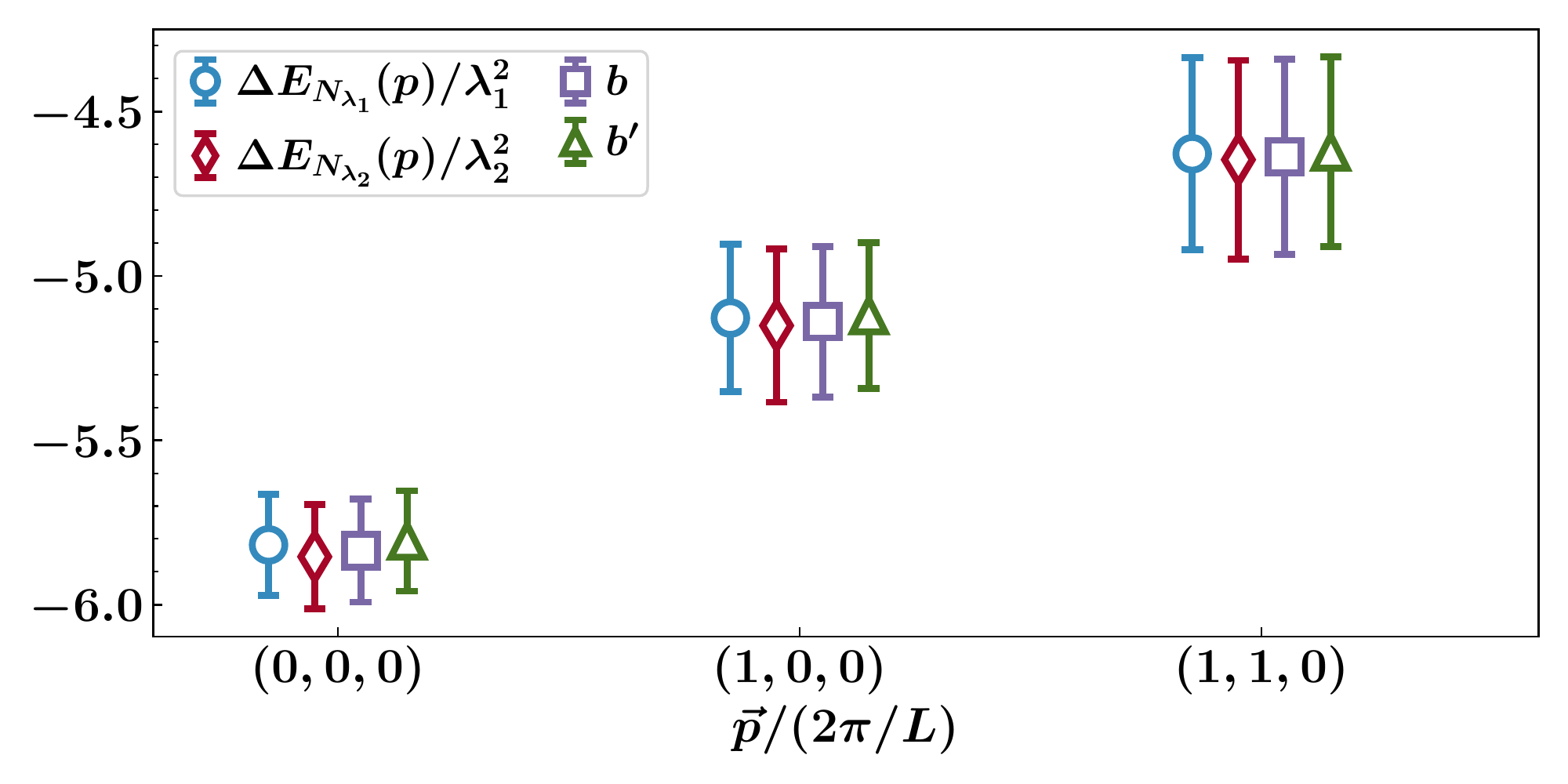}
	\caption{\label{fig:lambda_sys} The coefficient of the quadratic term in \Cref{eq:lamfit} determined in four different ways (see text). $b$ and $b^\prime$ are the quadratic coefficients obtained from a purely quadratic, $f(\lambda) = b \lambda^2$, and a quadratic-plus-quartic, $g(\lambda) = b^\prime \lambda^2 + c \lambda^4$, fit. We show the results for the $uu$ piece obtained on the $48^3 \times 96$ ensemble at fixed $\bv{q} = (5,3,0) \, \latmom$.
	 }
\end{figure*}

Fits to the energy shifts (\Cref{eq:lamfit}) are shown on the bottom rows of \Cref{fig:effmass_lamfit_1,fig:effmass_lamfit_2,fig:effmass_lamfit_3} for the $uu$ and $dd$ pieces of $T_{33}$ and $T_{00} + T_{33}$ both. Since the energy shifts at different $\lambda$ values are highly correlated, a $\chi^2$-based analysis is not a reliable goodness-of-fit test. However, we confirm the suppression of the $\mathcal{O}(\lambda^4)$ term, and the absence of $\lambda$-odd terms, by including $\mathcal{O}(\lambda)$, $\mathcal{O}(\lambda^3)$, and $\mathcal{O}(\lambda^4)$ terms separately in the fit. We find that the coefficient of the linear term is consistent with zero and any residual contamination from higher-order terms has a negligible effect compared to the statistical error on the extracted amplitudes. We show the coefficient of the quadratic term (\Cref{eq:lamfit}) for several nucleon momenta in \Cref{fig:lambda_sys} as determined in four different ways. We either normalise the energy shifts at each $\lambda$, $\Delta E_{N_{\lambda_i}} / \lambda_i^2$, or perform fits of the form $f(\lambda) = b \lambda^2$, and $g(\lambda) = b^\prime \lambda^2 + c \lambda^4$ that includes the quartic contamination. The data are well described by a purely quadratic fit, $f(\lambda)$, and any quartic contamination is negligible.

\Cref{fig:effmass_lamfit_2,fig:effmass_lamfit_3} compare the quality of the correlator ratios of $\omega = 0.06$ and $\omega=0.18$ regarding the discussion in \Cref{sec:res}. Although they lie close to each other in $\omega$ space (see \Cref{fig:F12L}), $\omega = 0.06$ has a larger nucleon momentum $\bv{p} = (-1,2,0) \, \latmom$, hence a worse S/N, leading to a larger uncertainty in the extracted amplitude as compared to the amplitude obtained for $\omega = 0.18$ $(\bv{p} = (0,1,0) \, \latmom)$. 

\begin{figure*}[ht]
	\centering
	\includegraphics[width=.6\textwidth]{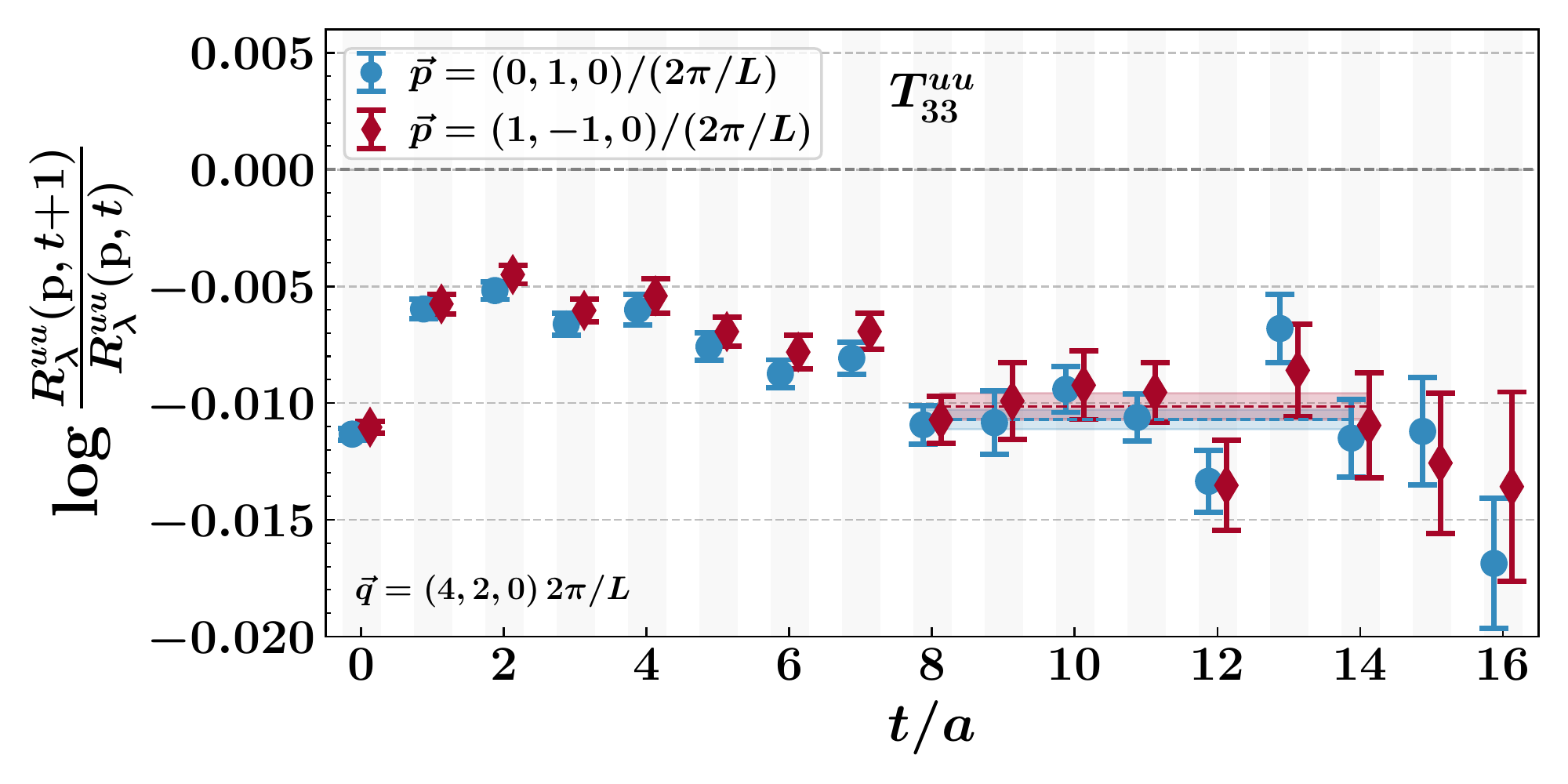} \\
	\includegraphics[width=.6\textwidth]{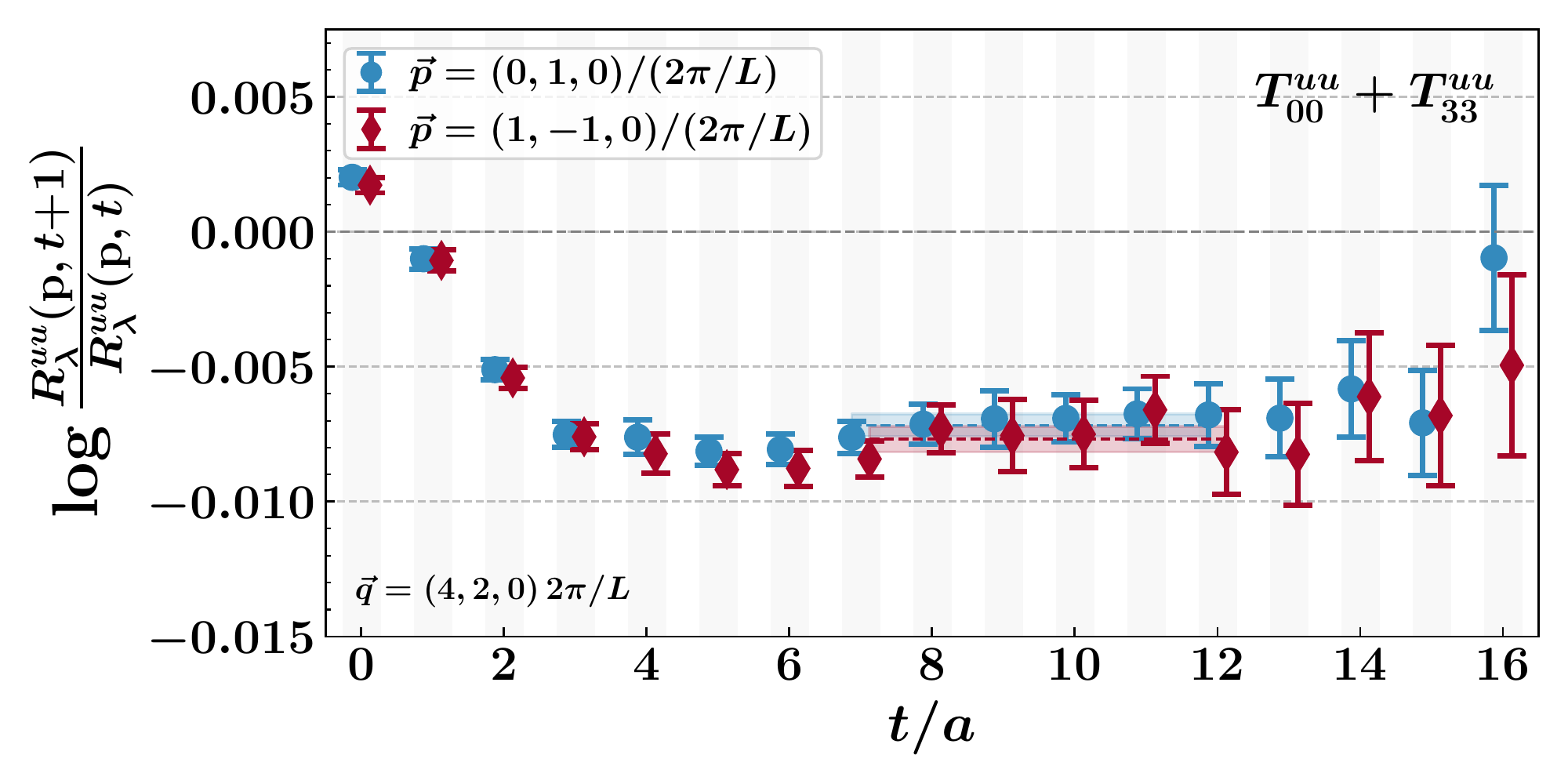} \\
	\includegraphics[width=.6\textwidth]{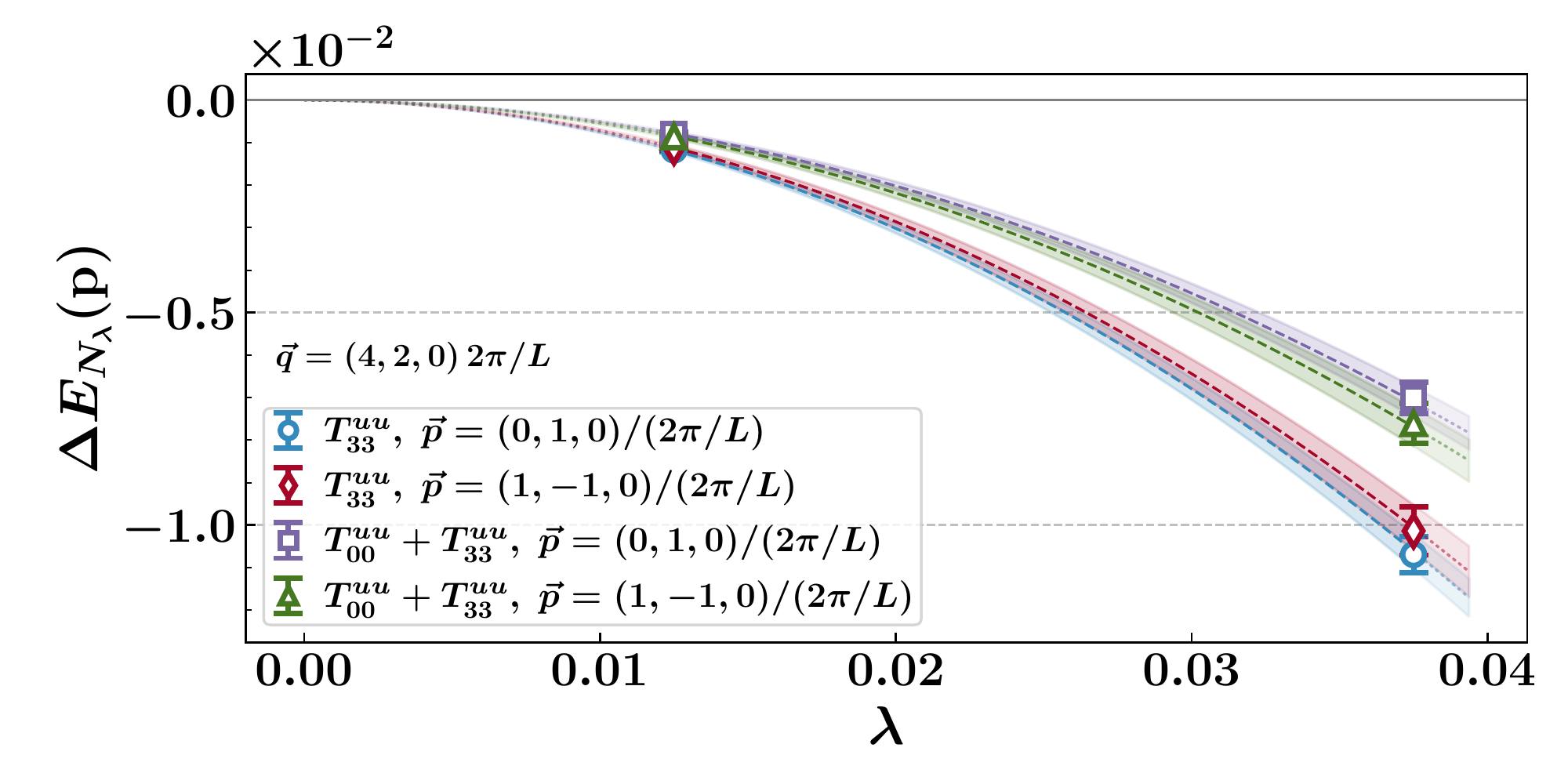}
	\caption{\label{fig:effmass_lamfit_same_w} Top to bottom: Correlator ratios of the amplitudes $T_{33}$ and $T_{00}+T_{33}$ for $\lambda = 0.0375$, and the corresponding fits in $\lambda$-space, respectively, for two different $\bv{p}$ momenta that give $\omega=0.2$ at fixed $\bv{q} = (4,2,0) \, \latmom$. See the caption of \Cref{fig:effmass_lamfit_1} for the explanation of shaded regions. We show the results for the $uu$ piece obtained on the $48^3 \times 96$ ensemble.}
\end{figure*}

A few of the $(\bv{p},\bv{q})$ pairs lead to the same $\omega$ for the kinematics considered in this work. We show the correlator ratios and fits to the extracted energy shifts for a representative case in \Cref{fig:effmass_lamfit_same_w} for the $(\bv{p}, \bv{q}) = ((0,1,0), \, (4,2,0)) \, \latmom$ and $((1,-1,0), \, (4,2,0)) \, \latmom$ pairs corresponding to $\omega=0.2$. We do not find any statistically significant deviation between the amplitudes extracted from such pairs and keep all occurrences if it is not omitted due to poor signal quality.    


\section{Bayesian analysis} \label{app:bayes}
We apply the same methodology employed in Ref.~\cite{PhysRevD.102.114505} to extract the moments of structure functions from our Compton amplitude data. Lowest non-vanishing moments, $M_2^{(1)}(Q^2)$ (\Cref{eq:moments1}) and $M_{0,2}^{(L)}(Q^2)$ (\Cref{eq:moments2}), are sampled from separate uniform distributions with bounds $[0,1]$, while the consecutive higher-moments are bounded from above by their respective preceding moment, $M^{(1,L)}_{2n}(Q^2) \in [0, M^{(1,L)}_{2n-2}(Q^2)]$, for $n > 1$. Bounds for the $ud$ moments are discussed in \Cref{sec:simu}. We employ the \texttt{PyMC} package, a probabilistic programming library for python~\cite{Salvatier:2016swf}, in our analysis.

\begin{figure*}[ht]
    \centering
    \includegraphics[width=.6\textwidth]{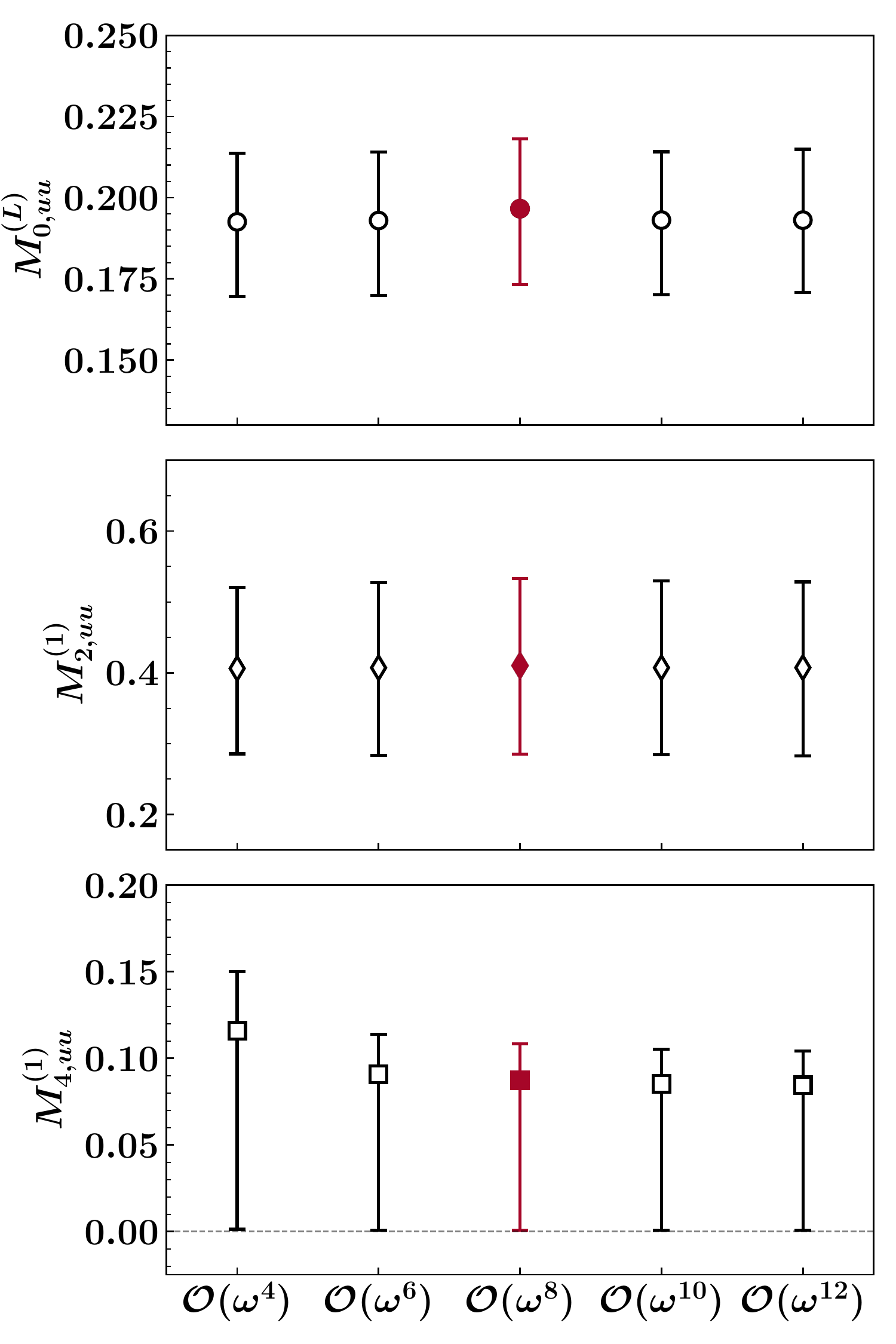} 
    \caption{\label{fig:stability} Stability plots for the lowest moments obtained on the $48^3 \times 96$ ensemble at $Q^2 = 4.86 \; {\rm GeV^2}$ for the $uu$ contribution only. $dd$ and $ud$ contributions behave similarly. We show $M_0^{(L)}$ (top) that is directly proportional to the lowest moment of $F_2$, and the lowest two moments, $M_2^{(1)}$ (middle) and $M_4^{(1)}$ (bottom) of $F_1$ with respect to the number of terms kept in the fit polynomials. Colour filled symbols indicate the values that we pick.    
    }
\end{figure*}

We keep terms up to $\mathcal{O}(\omega^8)$ in the fit polynomials~\Cref{eq:ope_moments1,eq:moments_F2}. We find this to be the minimum required number of terms to reliably extract at least the lowest two moments from our Compton amplitude data while keeping the computational overhead low. Keeping fewer terms lead to an overestimation of the moments, while including higher-order terms have a negligible effect. We illustrate the stability of the lowest moments in \Cref{fig:stability} for a representative case.

\begin{figure*}[ht]
    \centering
    \includegraphics[width=.6\textwidth]{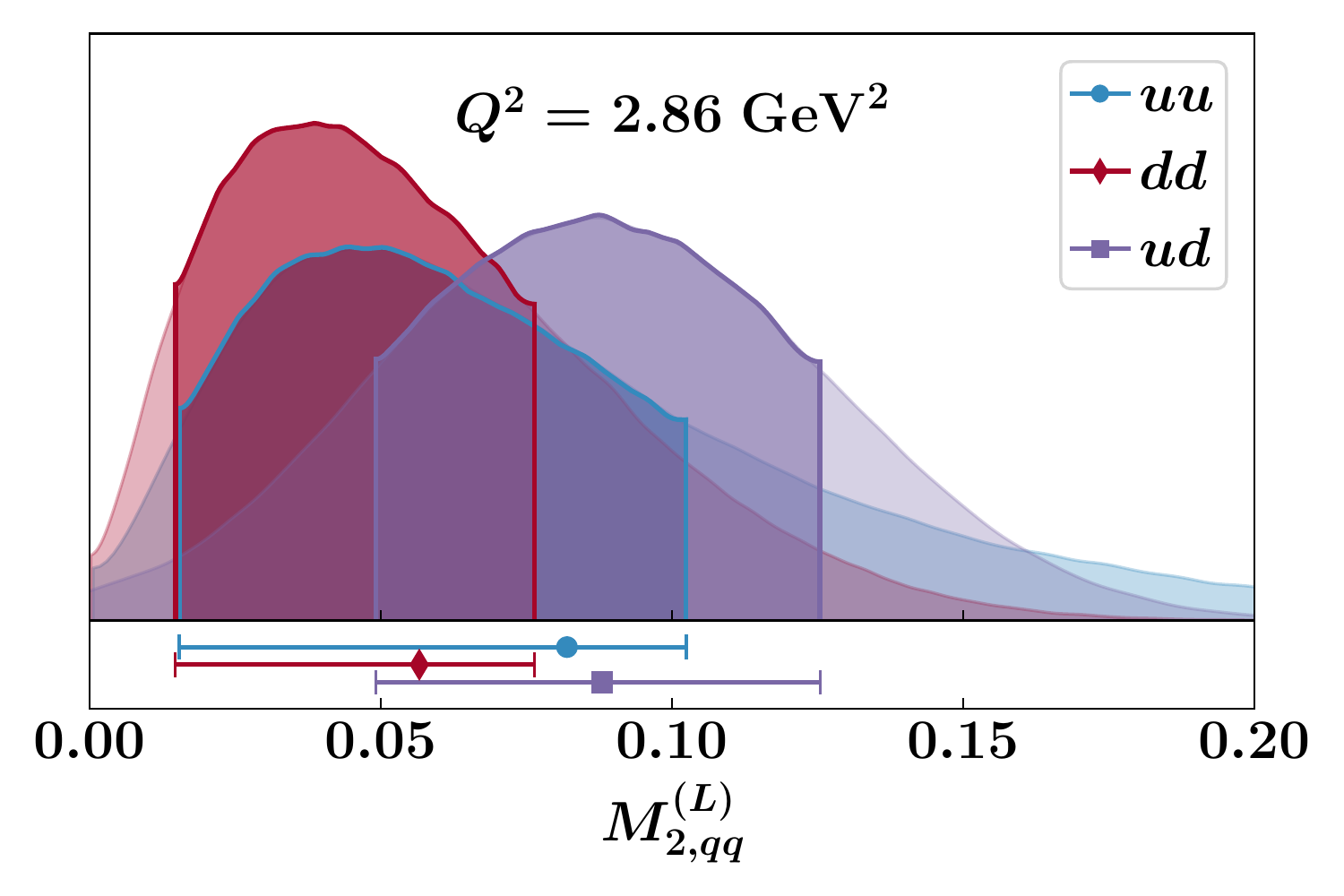}
    \caption{\label{fig:bayes_qq} Density plots of the posterior distributions for the lowest $M_{2}^{(L)}$ moments for $uu$, $dd$, and $ud$ contributions. We show the results for $Q^2 = 2.86 \; {\rm GeV^2}$ from the $48^3 \times 96$ ensemble. $68\%$ credible regions of the highest posterior density are indicated by the darker regions on the density plot and shown on the lower panel along with the means of the distributions.
    }
\end{figure*}

In \Cref{fig:bayes_qq} we show the inferred posterior distributions for the $M_{2}^{(L)}(Q^2)$ moments at $Q^2 = 2.86 \, {\rm GeV}^2$ for the $uu$, $dd$, and $ud$ contributions. Although the distributions of the $uu$ and $dd$ pieces are skewed towards zero, a non-zero signal is obtained for both. The $M_{2}^{(1)}(Q^2)$ and $M_{0}^{(L)}(Q^2)$ distributions (not shown) have well-defined Gaussian shapes. 

The lowest moments of proton $F_2$ and $F_L$ shown in \Cref{fig:F2_proton_moments,fig:FL_proton_moments}, respectively, are constructed using the individual $uu$, $dd$, and $ud$ contributions, $M_{i,p}^{(L)} = \frac{4}{9} M_{i,uu}^{(L)} + \frac{1}{9} M_{i,dd}^{(L)} - \frac{2}{9} M_{i,ud}^{(L)}$, where $i = 0, 2$. Given that $M_{2,uu}^{(L)}$ and $M_{2,dd}^{(L)}$ are skewed towards zero and having a $M_{2,ud}^{(L)}$ contribution as significant as $M_{2,uu}^{(L)}$, the resulting $M_{2,p}^{(L)}(Q^2)$ are highly-skewed towards zero making a clear exclusion of a zero value doubtful. Hence we are only confident in setting an upper bound for the $M_{2,p}^{(L)}(Q^2)$ moments. The $M_{0}^{(L)}(Q^2)$ moments, on the other hand, are directly proportional to the lowest moments of $F_2$, i.e. the intercepts of $\mathcal{F}_2/\omega$ shown in \Cref{fig:F12L}, and finite.

\end{document}